\documentclass[lettersize,journal]{IEEEtran}
\usepackage{amsmath,amsfonts}
\usepackage{algorithmic}
\usepackage{algorithm}
\usepackage{array}
\usepackage[caption=false,font=normalsize,labelfont=sf,textfont=sf]{subfig}
\usepackage{textcomp}
\usepackage{stfloats}
\usepackage{url}
\usepackage{verbatim}
\usepackage{graphicx}
\usepackage{cite}
\usepackage{siunitx}
\usepackage{booktabs}
\usepackage{bm}
\usepackage{enumerate}
\usepackage{hyperref}
\usepackage{tikz}
\usetikzlibrary{shapes,arrows.meta,positioning,fit}

\begin{document}

\renewrobustcmd{\bfseries}{\fontseries{b}\selectfont}
\sisetup{detect-weight,mode=text,group-minimum-digits = 4}

\def\z{\mathbf{z}}
\def\c{\mathbf{c}}
\def\q{\mathbf{q}}
\def\p{\mathbf{p}}

\def\X{\mathbf{X}}
\def\Z{\mathbf{Z}}
\def\C{\mathbf{C}}
\def\Q{\mathbf{Q}}
\def\P{\mathbf{P}}
\def\M{\mathbf{M}}

\def\sim{s}

\def\sec#1{Section~\ref{sec:#1}}
\def\fig#1{Figure~\ref{fig:#1}}
\def\tab#1{Table~\ref{tab:#1}}

\title{The Effect of Batch Size on Contrastive Self-Supervised Speech Representation Learning}

\author{Nik Vaessen and David A. van Leeuwen
\thanks{The authors are located at the Institute of Computing and Information Science, Radboud University, Nijmegen, The Netherlands. Contact: \{nvaessen,dvanleeuwen\}@science.ru.nl}
}



\maketitle

\begin{abstract}
Foundation models in speech are often trained using many GPUs, which implicitly leads to large effective batch sizes.  In this paper we study the effect of batch size on pre-training, both in terms of statistics that can be monitored during training, and in the effect on the performance of a downstream fine-tuning task. By using batch sizes varying from 87.5 seconds to 80 minutes of speech we show that, for a fixed amount of iterations, larger batch sizes result in better pre-trained models. However, there is lower limit for stability, and an upper limit for effectiveness. We then show that the quality of the pre-trained model depends mainly on the amount of speech data seen during training, i.e., on the product of batch size and number of iterations.  All results are produced with an independent implementation of the wav2vec 2.0 architecture, which to a large extent reproduces the results of the original work~\cite{wav2vec2_2020}. Our extensions can help researchers choose effective operating conditions when studying self-supervised learning in speech, and hints towards benchmarking self-supervision with a fixed amount of seen data. Code and model checkpoints are available at \url{https://github.com/nikvaessen/w2v2-batch-size}
\end{abstract}


\section{Introduction}
\IEEEPARstart F{oundation} models have become the norm in deep learning research. 
In the audio domain, popular models include wav2vec 2.0\cite{wav2vec2_2020,conneau_unsupervised_2021}, HuBERT\cite{hubert_2021}, and WavLM\cite{wavlm_2022}, whose model weights are generally available. 
These transformer models all use a form of self-supervised learning (SSL) with the use of a pretext task to learn (``pre-train'') speech representations. 
The models can then be fine-tuned on a myriad of downstream tasks\cite{yang21c_superb}, including speech recognition, speaker recognition, emotion recognition, and intent classification. 
However, self-supervised pre-training takes a tremendous amount of resources, exceeding high-end consumer grade hardware at the time of writing.
First, due to the unlabeled nature of self-supervision, it is relatively cheap to increase the dataset size. 
Over a span of two years, we have seen public training datasets increase by two orders of magnitude%
\footnote{Leaving aside Whisper\cite{whisper-radford23a} and Google USM\cite{usm_zhang_google_2023},with respectively 680\,k hours and with 12\,M hours of private training datasets.}%
, with wav2vec 2.0 using 1\,k hours of audio (circa 100\,GB) from Librispeech~\cite{panayotov_librispeech_2015}, to WavLM using 94\,k hours (circa 10\,TB) by combining Libri-light~\cite{kahn2020libri}, GigaSpeech~\cite{chen21o_gigaspeech} and VoxPopuli~\cite{wang-etal-2021-voxpopuli}.
Secondly, the seminal works mentioned above all report results of models trained with large batch sizes using data parallelism across many GPUs. 
For example, for models with 94\,M parameters, WavLM and HuBERT use 32 GPUs, and wav2vec 2.0 uses 64 GPUs, leading to batch sizes of respectively 45 and 90 minutes of audio.
The number of GPUs needed to work with these large batches in a timely manner, as well as the required disk space for the datasets, make it non-trivial to apply these algorithms.

While the effect of dataset size on performance is (at least partially) known\cite{wav2vec2_2020,conneau_unsupervised_2021}, to our knowledge there are no studies on the scaling behaviour of SSL algorithms with respect to the batch size and number of training iterations.
This can be of interest to researchers who do not have the resources to study these algorithms under large batch size conditions, or practitioners who need to make a trade-off between time, computational budget, and desired performance.
Given a fixed model complexity, dataset size, and number of training iterations, how much is gained by increasing the batch size? 
How well do these techniques work with fewer resources, and can the academic community do meaningful experiments without industrial-scale data centers?
While we aim to answer these questions generally, for precisely the reason of available computational resources, we limit ourselves to a single model and self-supervision algorithm, namely wav2vec 2.0\cite{wav2vec2_2020}. 
Concretely, we set out to address the following research questions:

\begin{enumerate}[{RQ }1{:}]
    \item \label{rq-bs-pre} How does the batch size affect the pre-training procedure of wav2vec 2.0? 

    \item \label{rq-bs-down} How does the batch size during pre-training affect downstream fine-tuning? 

    
    \item \label{rq-equiv} Can we compensate a reduction of the batch size by increasing the amount of training iterations by the same factor?

    
\end{enumerate}

Regarding all three RQs, and given the existing literature on speech SSL~\cite{wav2vec2_2020,hubert_2021,wavlm_2022}, our hypothesis is that large batch sizes are essential for pre-training convergence and the model's ability to be properly fine-tuned to the downstream task.   
For RQ \ref{rq-bs-pre}, we are interested in knowing whether a large batch size is a necessity for optimizing the objective.
It would be valuable to know the smallest possible converging batch size, and how optimization behaves with this batch size compared to the canonical, large batch size. For RQ \ref{rq-bs-down}, we expect that a larger batch size will lead to better downstream task performance, but we are especially interested in how much the performance improves with each doubling of the batch size.
What is the minimum batch size at which we see that fine-tuning is possible, and how does this depend on the amount of data available for fine-tuning?
For RQ \ref{rq-equiv}, we are interested in knowing whether training twice as long with half the batch size results in the same performance. 
This would imply that performance is only a function of how much data is seen during self-supervision, and that with patience, people with fewer resources can also carry out pre-training. 

To answer these questions, we pre-train wav2vec 2.0 with batch sizes ranging from 87.5 seconds to 80 minutes, and fine-tune for speech recognition, with 10 minutes to 960 hours of labeled speech. Hereby, we make the following contributions:

\begin{enumerate}
    \item We perform a comprehensive study of the effect of batch size and amount of training iterations for pre-training wav2vec 2.0, helping practitioners to make trade-offs when deciding on downstreak task performance.
    \item We show that the most important factor for the downstream task performance is the amount of data seen during self-supervision, indicating that fixing the product of batch size and training iterations in a benchmark can be valuable.
    \item We detail all subtleties present in the implementation of wav2vec 2.0 pre-training that are essential for achieving convergence
    \item We publicize an independent implementation of the framework\footnote{\url{https://github.com/nikvaessen/w2v2-batch-size}}, with matching results for the base model compared to~\cite{wav2vec2_2020}. We also provide all model checkpoints created throughout this work\footnote{See the code repository for a link to the model checkpoints}.
\end{enumerate}

This rest of this article is structured as follows. First, we will cover related work in \sec{related}, including studies on batch sizes with stochastic gradient descent, (contrastive) SSL (in speech) and its scaling behaviour, and research on SSL with smaller budgets. Then, \sec{method} will explain wav2vec 2.0 pre-training and fine-tuning, followed by experimental setup and results in \sec{experiments}, and we will close with a discussion and conclusion in \sec{conclusion}.

\section{Related work} \label{sec:related}

\subsection{Stochastic gradient descent and large batch sizes}

The authors of \cite{mccandlish2018empirical} study the trade-off between time and computational resources when choosing a batch size.
On the one hand, it is argued that a small batch size leads to gradients dominated by noise, which is averaged out by consecutive update steps.
In this case, one might as well compute multiple batches in parallel and average out the noise before applying an update, which is equivalent to increasing the batch size using data parallelism.
On the other hand, it is argued that when a batch size is very large, the gradient estimate contains little noise, and therefore sampling two batches and averaging their gradient will not lead to a significantly better estimate.
In this case doubling the batch size does not serve a practical purpose anymore.
Thus, there is a critical batch size, the exact value varying for each task and domain, after which an increase in batch size has strongly diminishing returns.
This critical batch size can be predicted with a metric called the gradient noise scale, and this scale can change throughout the optimization procedure.  

Complementary to \cite{mccandlish2018empirical}, the authors of \cite{Shallue2018dataparallel} did a comprehensive study on how batch size affects generalization performance, across multiple combinations of datasets (5 vision, 2 text), neural network families (FC, simple CNN, ResNet, VGG, LSTM, Transformer) and optimizers (SGD, with momentum, with Nesterov momentum).
First, they experimentally confirm the existence of a critical batch size, as reasoned in \cite{mccandlish2018empirical}.
Under all conditions diminishing returns were observed above a certain (different per condition) batch size.
Secondly, the magnitude of this critical batch size depends on the dataset, neural network type, and optimizer, and no clear relationship is found. 
Thirdly, they also do not find a clear relationships between the batch size and hyperparameters (such as the learning rate and learning rate schedule).
Lastly, increasing the batch size does not hurt generalization performance, although larger batch sizes sometimes require more regularization to achieve the same performance as a lower batch sizes.

Moreover, in \cite{Shallue2018dataparallel} the following points of concern were raised about comparing the performance between different batch sizes.
First, it is common to tune hyperparameters on a specific batch size, after which heuristics are used to pick hyperparameters for other batch sizes.
This gives a systematic advantage to the batch size for which the hyperparameters were tuned.
Secondly, when considering an epoch budget, smaller batch sizes are favored, as they can perform more update steps.
Thirdly, when considering an iteration budget, large batch sizes are favored, as they can see more data samples within the same amount of iterations.
Fourthly, noise in the gradients of small batches has a regularization effect, which needs to be corrected for with other regularization methods when using large batch sizes.
Lastly, the space of effective hyperparameters becomes smaller for large batch sizes. Therefore, they might require a more comprehensive hyperparameter search compared to small batch sizes.

\subsection{Self-supervised speech representation learning}

Representation learning concerns itself with being able to encode information in such a way that it makes learning a subsequent downstream task straightforward \cite{goodfellow2016deep}. 
For speech, a good representation would allow, e.g., phonemes to be linearly separable for a speech recognition softmax classifier, or speaker attributes to be distinctly clustered, so that a distance metric can be used to recognize speakers. 
Models which are trained in a supervised fashion already learn task-specific representations, as usually the output of a penultimate layer is used for classification purposes. 
In self-supervised representation learning, a pretext task is used instead, with the hope that solving this task requires learning representations which are also helpful for learning the actual downstream task(s) of interest.
These pretext tasks are designed such that they use some property of the input data itself as a label. 
This allows using much larger, and cheaper to collect, datasets, and for representations to potentially be useful for multiple distinct processing tasks. 
A good overview of pretext tasks used for speech representation learning is given in \cite{mohamed2022self}. 
They define three categories of pretext tasks, namely \emph{reconstructive}, \emph{contrastive}, and \textit{predictive}. 

Reconstructive pretext tasks limit the view of the speech signal, whereafter the model needs to fill in or complete the signal in some fashion.
Models using a reconstructive pretext task include VQ-VAE\cite{van2017neural}, Mockingjay\cite{liu2020mockingjay}, DeCoAR\cite{ling2020deep,ling2020decoar2}, and TERA\cite{liu2021tera}.
Later, it was argued in \cite{mohamed2022self} that reconstruction of speech results in heavily entangled representations, which makes them less useful for downstream tasks. 

Contrastive pretext tasks also limit the view of the speech signal.
However, instead of simply reconstructing, the objective focuses on predicting what information should or should not be encoded at a certain (unseen) time step of the signal.
This is done with an anchor representation, for which a target representation (which should be there) is distinguished from distractor representations (which should \textit{not} be there), thus creating contrast between learned representations.
Wav2vec\,2.0 is a popular contrastive model, which this work focuses on exploring more deeply.
Other contrastive models include Unspeech\cite{milde18_interspeech}, CPC\cite{oord2018representation}, and other wav2vec variants\cite{schneider19_interspeech,baevskivqwav2vec,sadhu21_interspeech,conneau_unsupervised_2021}.
The challenge of contrastive models, mentioned in \cite{mohamed2022self}, is the sampling of distractors.
For example, the learned representations can become invariant to speaker information if they are sampled from the same utterance.
Moreover, it is not clear how relatable the target and distractors are to the anchor, due to the difficulty of segmenting speech signals.   

Finally, predictive pretext tasks have known targets for the parts of the speech signal which were hidden.
It is these targets that the models learn to predict.
For example, HuBERT \cite{hubert_2021} uses cluster centroids of MFCCs computed over the training dataset as targets during self-supervision.
At one or more points during training, new targets are generated, by clustering the hidden representation outputs from the model, replacing the initial MFCC-based clusters.
Other examples of predictive approaches are WavLM\cite{wavlm_2022}, which is similar to HuBERT, but adds background speech to the input data which the model needs to learn to ignore, and data2vec\cite{baevski2022data2vec} as well as DinoSR\cite{liu2023dinosr}, which use a teacher-student approach, where targets are provided by a teacher model, which is updated with an exponentially moving average of the student model weights.
As argued in \cite{mohamed2022self}, the challenges of the mentioned models are mostly computational; HuBERT and WavLM require multiple iterations and good initial targets, whereas teacher-student approaches require twice as many model parameters during training.  

\subsection{Scaling self-supervised representation learning}

In this work we are interested in how contrastive speech SSL scales with the batch size and training duration.
In \cite{kaplan2020scaling} an analysis is made of the scaling behavior of pre-training large language models with respect to model size, dataset size, and the number of training steps.
Their primary finding is that the test loss follows a power law in relation to all three aspects, as long as model size is increased according to the dataset size, and training length is not made a bottleneck.
It is also found that very large models are more sample efficient, i.e., fewer iteration or less data is required compared to smaller models. 
A last finding is that model width and model depth does not really matter (within reason), as long as the number of parameters is increased, performance improves. 
In \cite{goyal2019scaling}, the scaling behavior of visual representation learning is studied, with respect to model size, dataset size, and complexity of the pretext task. 
They find that increasing the dataset size and complexity of the pretext is beneficial, as long as the model size is large enough.   
For speech SSL, the scaling behavior of the model size and the fine-tuning dataset size is studied in \cite{pu21_interspeech}.
They use the reconstructive pretext task Mockingjay\cite{liu2020mockingjay}, and their results match \cite{kaplan2020scaling}; larger model size leads to better performance, and larger models require less fine-tuning data.     

\subsection{Contrastive learning and batch size}

Contrastive self-supervision benefits from large batch sizes, as ablated in SimCLR\cite{pmlr-v119-chen20j}, and shown by, e.g., CLIP\cite{pmlr-v139-radford21a} and Florence\cite{yuan2021florence}. 
A hypothesis for this observation is that distractors are often sampled within the same mini-batch, and thus more (and potentially better) distractors are available as the batch size increases.
However, in \cite{pmlr-v137-mitrovic20a}, it is shown that computing the contrastive objective with fewer (e.g., only two) distractors per anchor leads to better performance, indicating that large batch sizes are the key factor of improved performance, and not the amount of available negative samples.  
In \cite{Chen2022ClGradBias}, they argue that small batch sizes in contrastive learning suffer from a gradient bias, which large batches sizes alleviate. 
Note that in wav2vec 2.0, negative samples are only taken from the same utterance. The batch size does not have any effect on the quality and quantity of negative samples, so there might be a gradient bias even with large batch sizes.  

\subsection{Self-supervised learning with academic budget}

The apparent effectiveness of large batch sizes makes it difficult to do research without a large computational budget. 
There has been some work on trying to reduce the resources required to do pre-training.
For example, in \cite{izsak-etal-2021-train} a BERT model is pre-trained to nearly equivalent performance with only 8 GPUs (with 12 GB VRAM) in 1 day, compared to 16 TPUs (with 32 GB RAM) and 4 days in the original work~\cite{devlin-etal-2019-bert}. 
This was done by reducing the maximum sequence length, focusing on large models, pre-masking data, and using specialised software packages such as DeepSpeed\cite{rajbhandari2020zero} and Apex\cite{MicikeviciusNAD18}.
Similar work has been done for the HuBERT model in \cite{chen23l_interspeech}. 
They show that using target representations from a fine-tuned ASR model in first iteration of HuBERT training (instead of MFCCs) leads to better performance while needing fewer GPU hours. 
Another line of thinking is presented in \cite{S3L}, where it is shown that self-supervised learning in vision can be done on small datasets, with low resolution images, and with models with relatively few parameters. 

\section{Methodology} \label{sec:method}

\begin{figure}
    \centering
    \tikzstyle{container} = [rectangle, draw, dashed]    
    \begin{tikzpicture}[>=Stealth, node distance = 1.1cm and 2cm, on grid, auto]
        \node (X) {$\X$};
        \node [
            rectangle,
            below=of X,
            rectangle split,
            rectangle split parts=2,
            draw,
            ] (CNN) {7-layer CNN \nodepart{two} grad scale};
        \node [below=1.25cm and 2cm of CNN] (Z) {$\Z$};
        \node [rectangle, draw, below right=of Z] (classify) {classify codewords};
        \node [below=of classify] (P) {$\P$};
        \node [rectangle, draw, below=of P] (quantize) {select codewords};
        \node [below=of quantize] (Q) {$\Q$};
        \node [below=of Q] (Qprime) {$\Q'$};
        \node [rectangle, draw, below left=of Z] (norm) {LayerNorm};
        \node [below=of norm] (Zprime) {$\Z'$};
        \node [rectangle, draw, below=of Zprime] (mask) {mask};
        \node [right=of mask] (M) {$\M$};
        \node [below=of mask] (Zhatprime) {$\hat\Z'$};
        \node [rectangle, draw, left=of Zhatprime] (rpe) {rel. pos. emb.};
        \node [circle, draw, below=of Zhatprime] (add) {+};
        \node [rectangle, draw, below=of add] (transformer) {encoder-only transformer};
        \node [below=of transformer] (C) {$\C$};
        \node [right=of Z] (penalty) {$\mathcal{L}_p$};
        \node [left=of P] (diversity) {$\mathcal{L}_d$};
        \node [right=of C] (Cprime) {$\C'$};
        \node [right =of Cprime] (contrastive) {$\mathcal{L}_c$};

        \node[container, minimum height=9.5em, fit=(rpe) (add) (transformer) (Zhatprime), label={95:context network}] (container) {};
        \node[container, fit=(X) (CNN), label={90:feature encoder}] (container) {};
    
    
        \draw[->] (X) -- (CNN);
        \draw[->] (CNN) -- (Z);
        \draw[->] (Z) -- (penalty);
    
        \draw[->] (Z) |- (classify.west);
        \draw[->] (classify) -- (P);
        \draw[->] (P) -- (quantize); 
        \draw[->] (P) -- (diversity);
        \draw[->] (quantize) -- (Q); 
    
        \draw[->] (Z) -| (norm.north);
        \draw[->,dashed] (norm) -- (Zprime); 
        \draw[->] (Zprime) -- (mask);
        \draw[->] (Zhatprime) -- (rpe);
        \draw[->] (mask) -- (Zhatprime);
        \draw[->] (mask) -- (M);
        \draw[->] (Zhatprime) -- (add);
        \draw[->] (rpe) -- (add);
        \draw[->] (add) -- (transformer);
        \draw[->] (transformer) -- (C);
    
        \draw[->,dashed] (C) -- (Cprime);
        \draw[->,dashed] (Q) -- (Qprime); 
        \draw[->] (Cprime) -- (contrastive);
        \draw[->] (Qprime) -- (contrastive);
        \draw[->] (M) -- (contrastive);
    \end{tikzpicture}
    \caption{A schematic overview of the wav2vec 2.0 framework during self-supervision. Dashed arrows indicate a projection using a linear layer without activation to match a target dimension. }
    \label{fig:schematic-w2v2}
\end{figure}
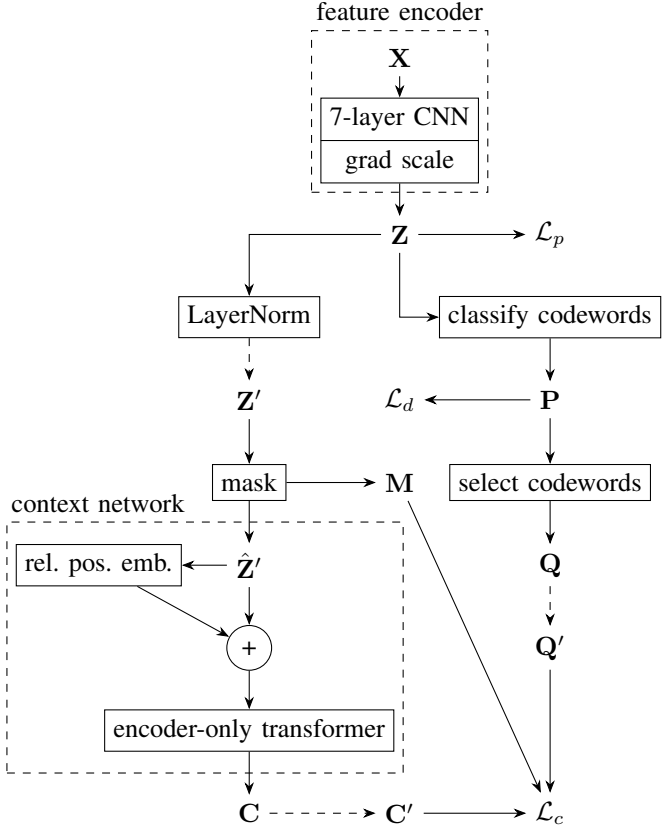

Here, we will decribe the architecture and pre-training procedure of wav2vec 2.0~\cite{wav2vec2_2020}, explicitly mentioning details we found to be essential for performance that received less attention in the original paper. We provide a schematic overview of wav2vec 2.0 in Figure \ref{fig:schematic-w2v2}, which can aid in understanding the dependence of different components and mathematical variables of the framework, which we will introduce below. 

\subsection{The CNN + Transformer network for audio}

In this work we use the standard architectural setup for self-supervised learning with audio \cite{wav2vec2_2020,hubert_2021, wavlm_2022}. 
First, the raw audio is processed into speech features with a 1-d convolutional neural network called the \textit{feature encoder}. 
This network has 7 convolutional layers, all of them with 512 channels.
For each respective layer, the kernel sizes are [10, 3, 3, 3, 3, 2, 2] with strides [5, 2, 2, 2, 2, 2, 2]. 
We include padding of [3, 1, 1, 1, 1, 0, 0] on both sides in order to change the output rate of the CNN to 50 feature vectors per second of audio, instead of slightly less as in \cite{wav2vec2_2020}. 
Each layer is followed by GELU activation, with tanh approximation. 
For the first convolutional layer, each output channel (thus, along the time dimension) is independently normalized to a learned mean and variance, by using GroupNorm \cite{wu2018groupnorm} with 512 groups and a learnable affine transform, before applying the activation function.
The CNN is followed by a gradient scaling layer, which during the forward pass acts as an identity function, but during the backward pass multiplies the global gradient with a constant. Hereby the magnitude of the gradients of the CNN can be controlled. The gradient scale constant is set to $\frac{1}{10}$ \cite{wav2vec2_2020}.
Mathematically, the feature encoder $f(\cdot)$ processes the raw 16\,kHz audio segment $\X = \{x_1, \dots, x_r\}$ to a sequence of latent speech feature vectors $\Z = \{ \z_1, \dots,  \z_T\}$, with $T=\lfloor r / 320\rfloor$.

The latent speech feature vectors are a local representation of the speech signal, in that each vector encapsulates speech factors in a window of 20\,ms.
A vanilla encoder-only transformer network called the \textit{context network} is used to create contextualized representations based on the local representations.
Due to self-attention in the transformer network, a contextual representation encapsulates speech factors from the entire audio segment. 
The transformer network has 12 layers, with an hidden dimension of 768 in the self-attention module, 12 attention heads, and a scale-up to 2048 dimensions in the feed-forward network, with GELU activation.
LayerNorm \cite{ba2016layernorm} is applied after the residual self-attention and feed-forward operations, encouraging the contextual representations to follow a multivariate normal distribution with a learned mean and variance.  

The local representations~$\z$ cannot directly be used as input to the transformer network.
First, LayerNorm is applied, so that the initial input features of the transform also follows a learned multivariate normal distribution. 
Then, a single linear layer (without activation) projects the local representation from 512 to 768 dimensions. 
The projected representations are then masked, which is described in more detail in section \ref{sec:mask}. 
The masking is an important aspect for pre-training, but is also beneficial during fine-tuning.
During inference no masking is done.  
As transformers need explicit positional information, a relative positional embedding is computed from the masked latent vectors, with a single convolutional layer, followed by GELU activation.
The convolutional layer has 768 output channels, a kernel size of 128, padding of 64 on both sides, and 16 groups. 
Moreover, weight normalization \cite{salimans2016weight} is applied on the kernel weights to aid convergence speed.
The stride is~1, therefore a relative positional embedding can be added to each latent vector.
Due to this summation, the input vectors of the transformer have a context of 1.25 seconds from both sides.

To regularize the network, dropout is applied on 3 locations%
\footnote{In~\cite{wav2vec2_2020} dropout is also applied on the latent speech features, after the projection layer. Additionally during SSL only, it is applied before quantization as well.  They also apply LayerDrop, which we skip to simplify the data-parallel implementation.}
in the transformer layers, namely on the self-attention scores (before weighted sum of values), on the output of the self-attention module (before residual addition), and on the output of the feed-forward network (before residual addition). 

To allow for independent modifications to the dimensions of each component in wav2vec 2.0, there are three so-called projection layers, consisting of a single fully-connected layer without activation.
In our notation we will use the accent~${}'$ to indicate a projected vector. 
The first projection layer is before the context network $g(\cdot)$, where the projection operates on \emph{normalized} $\z$. 
This $\z'$ is masked by a masking function $m(\z')$. 
Then, the context network processes the masked local representations $\hat\Z' = \{ \hat\z_1', \dots,  \hat\z_T'\}$ to contextual representations $\C = \{ \c_1, \dots,  \c_T\}$.  We will write $\C = g(\hat\Z')$ to indicate that each representation $\c_t$ had access to the entire sequence $\hat\Z'$.  The other locations where projection occurs is in the the computation of the contrastive loss, which we will describe below. 

\subsection{Self-supervision with contrastive learning}

In this work we study a contrastive learning approach to self-supervised speech representation learning called wav2vec~2.0\cite{wav2vec2_2020}.
Intuitively, the pretext task is to mask multiple regions of $\Z$, the local speech representation sequence, and feed this into the context network to predict the masked-out representations.
By enforcing contrast 
between predicted representations, the feature encoder has to place different information at different locations in the sequence $\Z$.
This can be seen as an intrinsic bias to learn phonetic units.
However, the context network does not directly predict masked $\z_t$ values.
Instead, each $\z_t$ is mapped to a quantized vector $\q_t$.
This $\q_t$ is part of a discrete set, and can be seen as a cluster centroid of the latent speech representation space, the cluster including $\z_t$.
It is this $\q_t$ the context network is trained to predict.
This is an additional intrinsic bias to learn phonetic units, as speech naturally clusters into acoustic units related to phonemes. 

We will follow with a detailed description of the quantization process, the masking strategy, and the objective function(s) of the pretext task.

\subsubsection{Quantization}

Each $\z_t$ in a sequence is individually classified to a quantized vector $\q_t$, thereby creating $\Q = \{\q_1, \dots, \q_T \}$.
The possible quantized vectors are learned, and represented by \textit{codebook}s, discrete sets of real-valued vectors.
A codebook $G$ is a set of $V$ entries (vectors) of a particular dimension $d_G$, representable as a matrix of size $V \times d_G$.
A single linear layer with softmax activation can be used to get a probability distribution over $V$ different classes.
The class with maximum probability can then be used to determine $\q_t$ from $\z_t$.
To (efficiently) increase the number of possible vectors, multiple codebooks and linear layers can be used, where the final quantized vector is the concatenation of the classification result from each codebook.
In wav2vec 2.0 there are two codebooks with each $V=320$ entries of size $d_G=128$, resulting in $320^2=102\,400$ quantized vectors with dimensionality $d_q=256$. 

However, the procedure above is not differentiable due to the selection.
This is circumvented by using the gumbel-softmax operation~\cite{jang2017categorical} after the linear layer, instead of using argmax on the softmax output.
The gumbel-softmax returns a one-hot logits vector during the forward pass, which implies that the discrete vector can be (differentially) selected with a weighted sum of the one-hot logits over the entries of $V$ in $G$. 
Additionally, like softmax, the gumbel-softmax can be controlled with a temperature parameter $\tau$.
At the start of training, $\tau=2$, which makes the gradient of codebook entries more uniform.
This is gradually decreased to $\tau=0.5$ during training, which makes the gradient of the non-selected codebook entries smaller.

\subsubsection{Masking} \label{sec:mask}

The masking is done in the context network, after the normalization and projection, but before computing the relative positional embedding.
The mask consists of multiple regions of $L_m = 10$ consecutive latent speech vectors which are all replaced by the same learned mask vector.
In total $p_m = 50\%$ of the latent vector sequence $\Z$ are masked, with the possibility that some regions overlap. 
The set of time steps where masking is applied is indicated by $\M$.
The number of mask regions for a given length $T$ is $n_r = \lfloor T p_m / L_m)\rfloor$. 
The length $T$ is excluding potential padding vectors from $T$ if present.  
The starting positions of the masked regions is determined by randomly choosing $n_r = \lfloor T/20\rfloor$ distinct time indices in the range $1, \ldots, T$.

\subsubsection{Objective function}

The objective function during SSL pre-training consists of a weighted sum of the main \textit{contrastive loss} $\mathcal{L}_c$, together with an auxiliary \textit{diversity loss }$\mathcal{L}_d$ with weighing $\lambda_d$, and an auxiliary \textit{L2 penalty loss} $\mathcal{L}_p$ with weighting $\lambda_p$:

\begin{equation}
\mathcal{L}_{\rm ssl} = \mathcal{L}_c + \lambda_d \mathcal{L}_d + \lambda_p \mathcal{L}_p    
\end{equation}

\textbf{Contrastive loss}. 
The contrastive loss encapsulates the pretext task, where we use the masked $\hat\Z'$ as input to the transformer in $g(\cdot)$, which has to predict the cluster centroids $\Q$ of the masked values in the output $\C$.
This prediction is done contrastively, such that for a given $\q_t$, the predicted $\c_t$ needs to be as similar as possible. 
At the same time, $\c_t$ needs to be as dissimilar as possible to time steps in $\Q \backslash \{\q_t\}$.
Dissimilarity is encouraged by sampling $k$ \emph{distractors} from $\Q$. 
The network is explicitly penalized if $\c_t$ is similar to any distractors sampled from $\Q$. 
Similarity is measured with the cosine similarity, written as $\sim(\mathbf{a}, \mathbf{b})$.
The loss can then be defined as

\begin{multline}
    \label{eq:loss_c}
    \mathcal{L}_c(\C, \Q, \M) = \\
    \sum_{t \in \M} -\log\Biggl(\frac
        {\exp\bigl(\sim(\c_t', \q_t') / \tau_c\bigr)}
        { \displaystyle \sum_{d\in D_t \cup \{t\}} \exp\bigl(\sim(\c_t', \q_d') / \tau_c\bigr)}
    \Biggr)    
\end{multline}
where $D_t$ is random sample of $k$ values from $\M\backslash \{t\}$.
Note that at any time the current time step $t$ is excluded from the sampling. 
The values $\c_t$ and $\q_t$, with dimensionality $d_c=768$ and $d_q=256$ are respectively projected to $\c_t'$ and $\q_t'$, both with dimension $d_{\rm sim}=256$ so that the cosine similarity can be computed. 
A temperature $\tau_c=0.1$ leads to a hard softmax distribution, controlling the gradient to focus on making correct predictions more than being dissimilar to distractors.
The contrastive loss can be interpreted as a standard $1+k$ classification task with the cross-entropy criterion. The logits are provided by $\sim$, softmax is applied over the logits, and the target is always the class index representing $\q_t$. 

\textbf{Diversity loss}.
A shortcut to optimizing the contrastive loss is to map all values in $\Z$ to the same quantized vector.
To prevent this, a diversity loss is applied, which encourages uniform predictions over the codebook entries.
A codebook $G$ with $V$ entries has classified the sequence $\Z$ to $\Q$, using logits from a softmax activations of a linear layer $\P = \{\p_1, \dots, \p_T\}$. For uniform predictions the average probability distribution $\bar{\p} = T^{-1}\sum_{t=1}^T \p_t$ should be flat. In this \emph{best} case, the entropy $H(\bar{\p}) = \log V$,  
and the perplexity $e^{H(\bar{\p})} = V$. In the case of \emph{shortcut}, a single class has probability 1, which means the entropy $H(\bar{\p})=0$ and the perplexity $1$. Therefore, the diversity loss minimizes the number of the entries in a codebook subtracted by the perplexity of the predictions:  
\begin{equation}
    \mathcal{L}_d(\P) = V -\exp\bigl(-\sum_{j=1}^S \bar{p}^{(j)} \log \bar{p}^{(j)} \bigr),
\end{equation}
where $\bar{p}^{(j)}$ is the $j$th component of $\bar\p$. 
Note that for the diversity loss, the logits of the linear layer are activated with vanilla softmax, while the selection of $\Q$ is done by activating the logits with the gumbel softmax (including a temperature $\tau$). In practise, there can be multiple codebooks, in our case we have $G_1$ and $G_2$, with equal number of entries~$V$ and dimension $d_G$. We compute the loss separately for both codebooks and sum the result. The weighting $\lambda_d = \frac{1}{10}$ throughout this work. 

\textbf{L2 penalty loss}. 
The third loss is a regularization term, which keeps the values of $\Z$ as small as possible. This loss is defined as
\begin{equation}
    \mathcal{L}_p(\Z) = \frac{1}{Td_z} \sum_{t=1}^T \sum_{j=1}^{d_z} \bigl(z_t^{(j)}\bigr)^2,
\end{equation}
where $z_t^{(j)}$ is the $j$th component of $\z_t$, and $d_z=512$.
The weighting $\lambda_p=10$ throughout this work.

\subsection{Batch creation}

So far, the descriptions have assumed a single utterance $\X$, while training is done with a batch of the dataset, split into multiple gpu-batches for distributed data-parallel training.
The LibriSpeech dataset is used, implying utterances have variable lengths, at minimum 0.83 seconds, and at most 30 seconds.
As each utterance in a gpu-batch needs to have the same length, all but the longest raw waveform in a batch are padded with zeros.
To minimize the amount of padding, the utterances are sorted by length, and put into bins of 5000 utterances.
Each gpu-batch is sampled from only a single bin.  Random samples from the bin feed a priority queue of length 50, from which gpu-batches are formed by taking samples prioritized by shortest duration, until the total speech duration in the gpu-batch exceeds a threshold. 
Because of limitation of the GPU memory, a gpu-batch is discarded if the difference between the shortest and longest utterance is more than 10 seconds.

The feature encoder network also processes the padded part of utterances. However, every vector $\z_t$ which results purely from padding in the raw waveform are ignored in the self-attention of the transformer by setting their attention score to $-\infty$. When creating the mask $\M$, the padded vectors $\z_t$ are also not considered part of the utterance. The contrastive and L2-penalty loss can be computed independently for each utterance in the gpu-batch, and summed afterwards. For the diversity loss, the probability distribution $\bar{\p}$ is computed by averaging over all predictions of all utterances in the gpu-batch, before computing the perplexity. The gradient resulting from each gpu-batch are averaged before the weights of the networks are updated. 

\subsection{Fine-tuning for speech recognition}

To fine-tune a pre-trained model, $\Z$ and $\C$ can be computed from $\X$, disregarding the quantization. The network still applies a mask, but only $p_m=5\%$ of the utterance is replaced with the learned masking vector. This acts as a regularization method, similar to SpecAugment~\cite{park19e_specaugment}. Each vector in $\C$ can be separately classified to a character (or blank) with a softmax-activated linear layer, and optimized with CTC~\cite{Graves:2006} loss. The feature encoder is generally not updated when fine-tuning, and the context network is only updated after the first 5\,k iterations.  





\section{Experiments} \label{sec:experiments}

\subsection{Pre-training with different batch sizes}

The first experiment aims to directly answer RQ~\ref{rq-bs-pre}, and is a prerequisite for answering all others. How does the batch size affect wav2vec2 pre-training?

\subsubsection{setup}

We pre-train the BASE wav2vec2 network with batch sizes ranging from 87.5 seconds to 80 minutes of audio, as seen in Table \ref{tab:batch_size_lr}.
Each pre-training starts with the same initial weights. 
We use all 960 hours of training data in LibriSpeech. 
We split off 5\% of each training subset (clean-100, clean-360, other-500) as a validation set. 
As self-supervision involves a lot of computational resources, we adhere to the hyperparameters as published in the seminal paper\cite{wav2vec2_2020} as much as possible. 
We use 400\,k training iterations with the AdamW optimizer, but change to a 8-cycle triangular learning rate where one cycle has 25\,k linear steps up and 25\,k linear steps down.  
The base learning rate of the cycle is 100 times smaller than the maximum learning rate.   
Using a cycling learning rate schedule makes it possible to directly compare different training lengths. 
This allows us to answer RQ~\ref{rq-equiv} with a single training run.
We also use GPUs with 24\,GB of VRAM, and therefore fill each GPU with a maximum of 2.4\,M audio samples (150 seconds) for full utilization of the device, compared to 1.4\,M samples (87.5 seconds).
This means that our experiment with 32 GPUs (batch size of 80 minutes) is close to the original experiment with 64 GPUs (batch size of 90 minutes) in \cite{wav2vec2_2020}.

For each batch size of duration $s$, we need to find a well-performing maximum learning rate (LR) for the cyclic schedule. 
As a full hyperparameter search would exceed our computational budget, we use heuristics to choose three different learning rates, and settle on a run with the lowest overall validation loss. 
We validate every 5\,k steps.
The first heuristic is to scale the learning rate linearly with the batch size. 
As a reference, a maximum learning rate of $m_{\text{lr}} = \num{5e-4}$ was used in \cite{wav2vec2_2020} together with a batch size of circa 1.6 hours. 
Therefore we use $h_{\text{lin}}(s) = m_{\text{lr}} s / s_{\text{orig}}$ as the first heuristic for the learning rate, with $s_{\text{orig}}=6000$ seconds%
\footnote{In~\cite{wav2vec2_2020} a batch size of 5600 seconds is used, but we use 6000 seconds for this heuristic calculation so that $h_{\text{lin}}$ rounds nicely. We still find well-performing LRs.}. 
The second heuristic is to scale the learning rate with the square root of the batch size. 
We use $h_{\text{sub}}(s) = m_{\text{lr}} \sqrt{s / s_{\text{orig}}} $ as the sub-linear learning rate heuristic.
For each batch size we also try the constant $h_{\text{const}}(s) = m_{\text{lr}}$, although this led to divergence for $s \leq 600$ seconds.

\begin{table}[t]
\centering
\caption{All attempted batch sizes for self-supervision, together with the number of required GPUs, and 3 possible learning rates. The bold learning rates resulted in the lowest validation loss. For the 32 GPU setting we only tried 1 learning rate.}
\label{tab:batch_size_lr}
\begin{tabular}{@{}rrrrrr@{}}
\toprule
\multicolumn{2}{c}{batch size}                    & \multicolumn{1}{c}{}     & \multicolumn{1}{c}{}         & \multicolumn{1}{c}{}                 & \multicolumn{1}{c}{}                  \\ \cmidrule(r){1-2}
\multicolumn{1}{c}{sec} & \multicolumn{1}{c}{min} & \multicolumn{1}{l}{GPUs} & \multicolumn{1}{c}{$h_{\text{const}}$} & \multicolumn{1}{c}{$h_{\text{sub}}$} & \multicolumn{1}{c}{$h_{\text{lin}}$} \\ \midrule
87.5                    & 1.5                     & 1                        & \num{5.00e-4}             & \num{6.04E-05}                             & \textbf{\num{7.29E-06}}                     \\
150                     & 2.5                     & 1                        & \num{5.00e-4}             & \textbf{\num{7.91E-05}}                    & \num{1.25E-05}                              \\
300                     & 5                       & 2                        & \num{5.00e-4}             & \textbf{\num{1.12E-04}}                    & \num{2.50E-05}                              \\
600                     & 10                      & 4                        & \num{5.00e-4}             & \textbf{\num{1.58E-04}}                    & \num{5.00E-05}                              \\
1200                    & 20                      & 8                        & \num{5.00e-4}             & \textbf{\num{2.24E-04}}                    & \num{1.00E-04}                              \\
2400                    & 40                      & 16                       & \textbf{\num{5.00e-4}}    & \num{3.16E-04}                             & \num{2.00E-04}                              \\
4800                    & 80                      & 32                       & \textbf{\num{5.00e-4}}    & \multicolumn{1}{c}{-}                      & \multicolumn{1}{c}{-}                       \\ \bottomrule
\end{tabular}
\end{table}

\subsubsection{results}

\begin{figure*}[t]
    \centering
    \includegraphics[width=\textwidth]{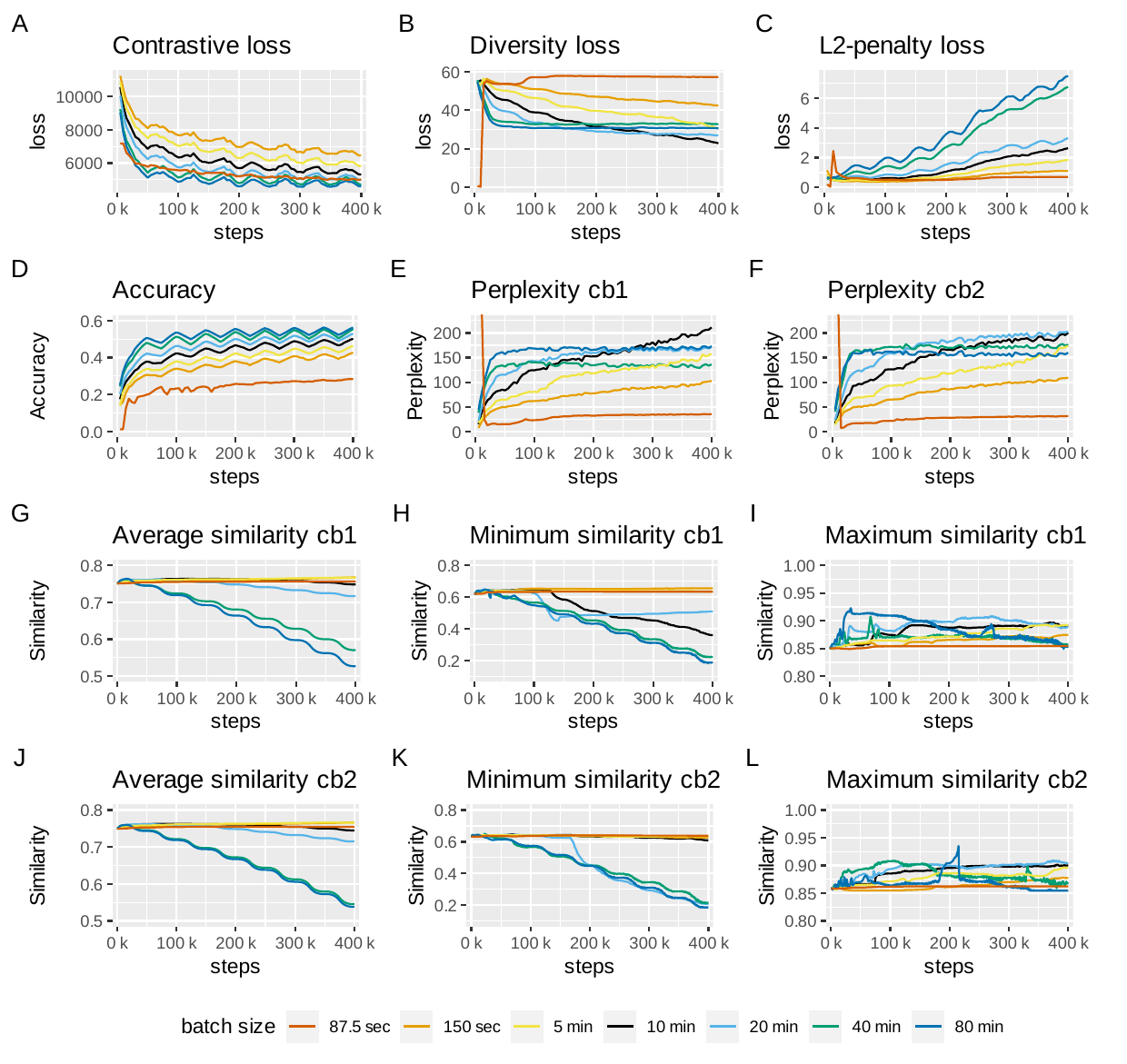}
    \caption{Various metrics on validation data (interval of 5\,k training steps) during self-supervised pre-training with different batch sizes, namely all three losses (A, B, C), the accuracy of predicting the correct masked quantized vector (D), and the perplexity of codebook 1 (E) and codebook 2 (F). We also show the average, minimum, and maximum value of the cosine similarity between codewords of codebook 1 (G, H, I) and codebook 2 (J,K,L), with an interval of 100 training steps.}
    \label{fig:ssl-metrics}
\end{figure*}

We show various metrics during the training procedure in \fig{ssl-metrics}.  
For the contrastive loss (A), we see that overall a larger batch size leads to a lower loss, except for the smallest batch size, 87.5 seconds.
For the diversity loss (B), we see that a large batch size (40, 80 min) causes the loss to drop quickly, but then plateau.
The other batch sizes steadily decrease, with a batch size of 10 minutes having the sharpest decrease. 
Notably, for batch sizes of 10 and 20 minutes the diversity loss surpassed the values of batch sizes 40 and 80 minutes after 150\,k to 200\,k steps.
The lowest batch size, 87.5 seconds, converged up to 5\,k steps, but then diverged immediately after.
The same patterns visible in the diversity loss are also seen in the perplexity of the codebooks (H and I, with the vertical scale reversed).
Without exceptions, for the L2-penalty loss (C), larger batch sizes lead to a higher loss. 
For accuracy (D), a larger batch size leads to higher accuracies, although there is a minimal difference between a batch size of 40 and 80 minutes.
For the similarity of codewords within the codebooks, we see that the average (G, J) only goes down steeply with large batch sizes (40 and 80 min).
This is similar for the the minimum observed similarity, with the additional that for batch sizes of 10 and 20 minutes we also observe lower minimum similarity over the training procedure.
Note that the inner product between any two codewords is always positive. 
For the maximum similarity values, we observe that they stay relatively stable, although for the larger batch sizes they increase slightly at the start of training, but decrease again during the training procedure, which can be related to the decay strategy of $\tau$ used in the gumbel-softmax. 

\subsection{ASR fine-tuning with varying amounts of labels} \label{sec:asr-ft}

The second experiment focuses on RQ \ref{rq-bs-down}. How is downstream fine-tuning affected by the batch size during pre-training?

\subsubsection{setup}

For each batch size in Table \ref{tab:batch_size_lr} we have (multiple) self-supervised training runs of 400\,k steps, with checkpoints saved every 5\,k steps.
For each batch size we select the run (and step) with the lowest overall validation loss, resulting in a single checkpoint which is used as initialization for training a speech recognition system.
This is the checkpoint at step 400\,k for all runs, expect for batch size of 32 GPUs (80 minutes), which had the lowest validation loss at step 305\,k.
For each of these selected checkpoints we perform a fine-tuning on 10 minutes, 1 hour, 10 hours, 100 hours, and 960h hours of labeled LibriSpeech data, as was done in \cite{wav2vec2_2020}.
For fine-tuning we use a tri-stage learning rate, with a base LR of \num{5e-7}, which linearly grows to \num{5e-5} in 10\% of the total iterations, then stays constants for 40\% of total iterations, and exponentially decays over the remaining steps to \num{2.5e-6}.
We use a total of 12k (10 min), 13k (1h), 20k (10h), 50k (100h) or 320k (960h) steps, with a batch size of 3.2\,M samples (200 seconds) on a single GPU.
For the first 5\,k steps the context network is frozen, and the feature encoder is frozen throughout the whole fine-tuning procedure.
We use a dropout of 10\% in the transformer network, and mask up to 5\% of the latent speech features vectors of each audio file in a batch.
We do not use LayerDrop.
We show results with greedy letter decoding, as well as with word decoding using a 4-gram LibriSpeech language model.
For word decoding we use use a beam size and threshold of~50, a language model weight of~2, and a word insertion score of~0 for all settings. 

\subsubsection{results}

\begin{figure*}[t]
    \centering
    \includegraphics[width=\textwidth]{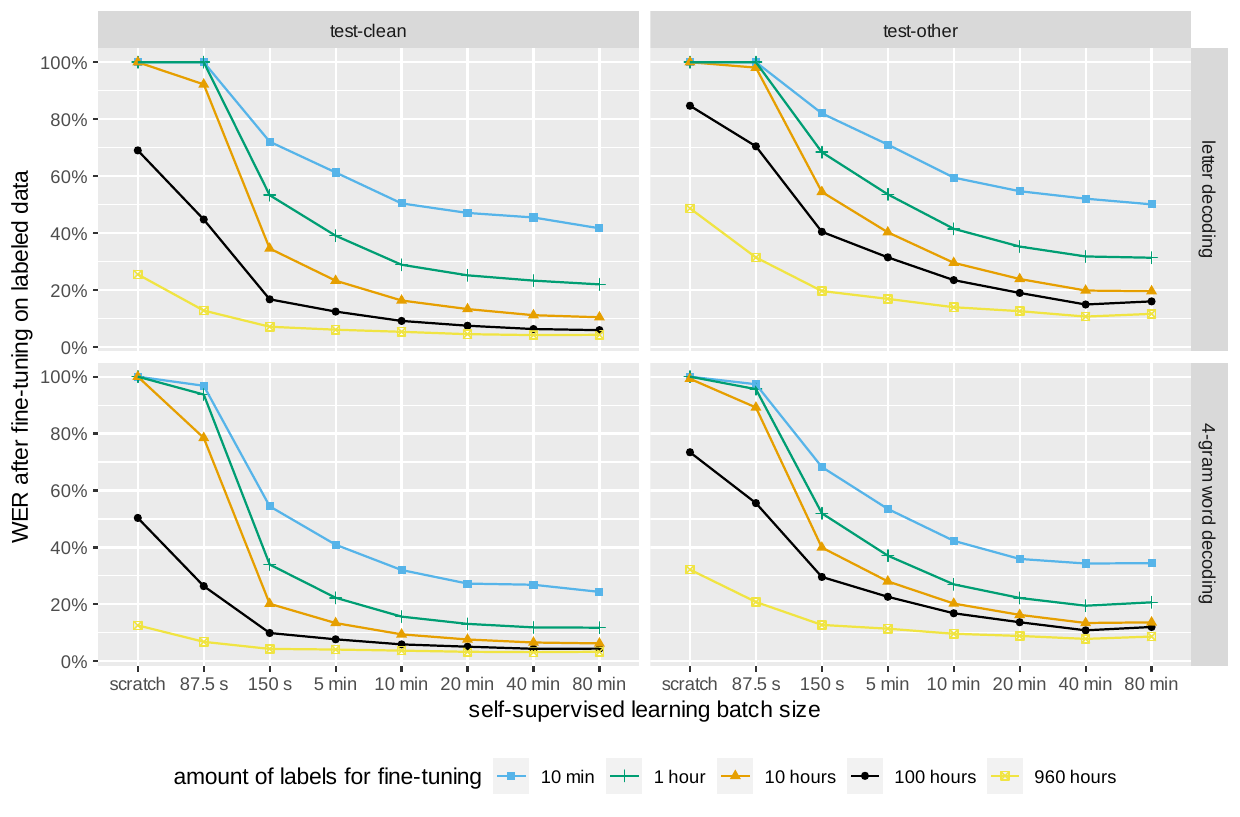}
    \caption{The WER (left column: librispeech test-clean, right column: librispeech test-other) against the batch size during pre-training of a self-supervised initialization. The self-supervised models are fine-tuned for speech recognition using 5 different magnitudes of labeled data. Scratch indicates fine-tuning a random initialization instead of a self-supervised initialization. The upper row shows the WER with letter decoding, while the bottom row shows the WER with word decoding using a 4-gram language model. }
    \label{fig:ft-asr-all}
\end{figure*}

We show the WER, evaluated on librispeech test-clean and test-dev, for each fine-tuning condition in \fig{ft-asr-all}. 
Two clear patterns are visible.
First, independent of the amount of fine-tuning labels available, we observe that fine-tuning a random initialization leads to the highest WER.
Then, each consecutive increase in the batch size during self-supervised learning leads to lower WERs. 
There is one exception: on test-other, the 40 min batch size initialization performs better than the 80 min batch size initialization, but only when fine-tuning with 10 or more hours of labeled data.
We observed similar degraded performance after fine-tuning the 400\,k checkpoint with batch size of 80 minutes (not shown in \fig{ft-asr-all}).
Secondly, having more labeled data for fine-tuning leads to a lower WER for each self-supervised batch size.
However, the larger the batch size, the smaller the difference in WER between the amount of labels available during fine-tuning.
Notably, \cite{wav2vec2_2020} reports 9\%/47\% WER on test-clean with/without a language model when fine-tuning with 10 minutes of labeled audio.
In this experiment we observe a WER of 24\%/41\% instead, the large difference in LM performance we attribute to our much smaller beam size in decoding.  
Finally, we see diminishing returns at a batch size of 80 min.

\subsection{Analysis on amount of data seen during self-supervision}

\begin{table}[]
\centering
\caption{The number of epochs and total amount of data observed throughout pre-training for each batch size. The training dataset contains 912 hours of data and we train for 400\,k iterations. Note that the batch size is an upper bound as they are constructed with variable length samples.}
\label{tab:epochs}
\begin{tabular}{@{}rrlcclcc@{}}
\toprule
\multicolumn{2}{c}{batch size} &  & \multicolumn{2}{c}{upper bound} &  & \multicolumn{2}{c}{measured} \\ \cmidrule(r){1-2} \cmidrule(lr){4-5} \cmidrule(l){7-8} 
\multicolumn{1}{c}{sec} & \multicolumn{1}{c}{min} &  & epochs & \multicolumn{1}{l}{\begin{tabular}[c]{@{}l@{}}observed \\ data (h)\end{tabular}} &  & \begin{tabular}[c]{@{}c@{}}max. repeats\\ of sample\end{tabular} & \begin{tabular}[c]{@{}c@{}}observed\\  data (h)\end{tabular} \\ \cmidrule(r){1-2} \cmidrule(lr){4-5} \cmidrule(l){7-8} 
87.5 & 1.5 &  & 11 & 10\,k &  & 11 & 9\,k \\
150 & 2.5 &  & 18 & 17\,k &  & 18 & 16\,k \\
300 & 5 &  & 37 & 33\,k &  & 36 & 31\,k \\
600 & 10 &  & 73 & 67\,k &  & 71 & 62\,k \\
1200 & 20 &  & 146 & 133\,k &  & 140 & 124\,k \\
2400 & 40 &  & 292 & 267\,k &  & 277 & 248\,k \\
4800 & 80 &  & 585 & 533\,k &  & 554 & 497\,k \\ \bottomrule
\end{tabular}
\end{table}

So far, we have observed that larger batch sizes lead to a lower contrastive loss, and less similarity between codewords.
We have also seen that larger batch result in better fine-tuning performance for speech recognition, irrespective of the amount of labels available.
Why are large batch sizes more effective?
Two hypotheses come to mind.

First, a large batch size better approximates the true gradient of the objective function as noise is averaged out better. 
Moreover, in contrastive learning, there is evidence that there is a gradient bias due to only seeing negative samples from the same batch\cite{Chen2022ClGradBias}.
A larger batch size can alleviate this bias as more negative samples are available.
We argue that for wav2vec 2.0 pre-training, a large batch size only allows for making more accurate optimisation steps by having less noisy gradient approximation.
This is because the negative examples are chosen only from within the masked region of the same utterance, and therefore the amount and quality of negatives does not scale with the larger (aggregated) batch size. 
Subsequently, having more accurate gradients implies you can use higher learning rates, and you can therefore better and more quickly approach desirable speech representations. 

Secondly, a large batch size allows the learning algorithm to observe more data with the same amount of iterations. 
This is shown in Table \ref{tab:epochs}. 
With the lowest batch size only 9\,k hours of data is observed, while using the largest batch size leads to observing almost 500\,k hours of data. 
Large batch sizes could simply be effective because it allows us to efficiently parallelize the learning algorithm. 
With our hardware the maximum batch size of a single GPU is 150 seconds and it takes 0.4 seconds to complete an iteration.
Observing 500\,k hours on a single GPU with this batch size would require 12\,M steps, and it would take 56 days. 
Instead, by using a large batch size and parrellizing across multiple GPUs, the runtime for observing 500\,k hours of data can be reduced significantly.
With a batch size of 80 minutes and 32 GPUs you need only 2 days%
\footnote{We used 8 GPUs with 4 accumulation steps for a run time of 8 days.}%
for the algorithm to process 500\,k hours of data.  

We will study these hypotheses with two experiments. One experiment will look at the variance of the gradients and how it relates to the batch size. The other experiment will compare downstream performance between pre-training conditions with the same amount of data seen, but using a different batch size and number of training iterations. 

\subsubsection{Variance of gradients}

First, we compare the gradients between different batch sizes. 
If the gradients are more accurate, and less noisy, with increased batch sizes, we expect the variance of the gradient to decrease.
To verify this we use the saved checkpoints (every 5\,k steps for 400\,k steps) during pre-training.
We restart pre-training with a checkpoint and the corresponding batch size, and compute the gradient approximations of 10 new, independently sampled batches from the training set.
For each iteration we store the resulting gradient vector of the backward pass, without taking into account the AdamW optimizer state nor the learning rate.
For each parameter, we compute the variance over the gradient approximations of the 10 batches. We can also compute the overall variance of the gradient vector by averaging over all parameters.
\begin{figure}
    \centering
    \includegraphics[width=0.5\textwidth]{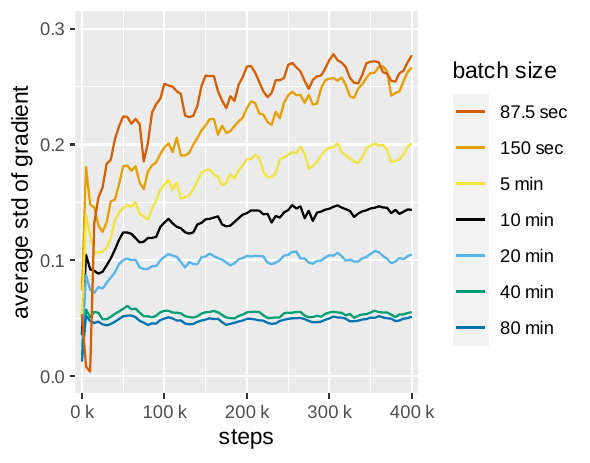}
    \caption{The standard deviation of the gradient, averaged over all parameters, against consecutive checkpoints, every 5\,k steps during pre-training. The gradients are calculated with 10 random batches, and the size of the batch matches the size used when training the checkpoint. No update steps are applied during these measurements.}
    \label{fig:grad-var}
\end{figure}
The average standard deviation against the training iterations for each batch size is shown in \fig{grad-var}. 
We observe that the standard deviation decreases as the batch size increases.
Further, the critical batch size is around 40 minutes; the standard deviation barely decreases when the batch size is doubled to 80 minutes.
Furthermore, for small batch sizes the standard deviation increases over the training procedure, while it stays constant for large batch sizes.
We also observe that the small batch sizes of 87.5 and 150 seconds have the same variance after 400\,k training steps.
Finally, the cyclic learning rate schedule seems to affect the gradient variance, as it is relatively high when the cyclic is at the minimum learning rate, and relatively low when at the maximum learning rate.
Note that these measurements were taken without considering the learning rate.

\subsubsection{Fine-tuning after seeing certain amounts of data}

\begin{figure*}[t]
    \centering
    \includegraphics[width=\textwidth]{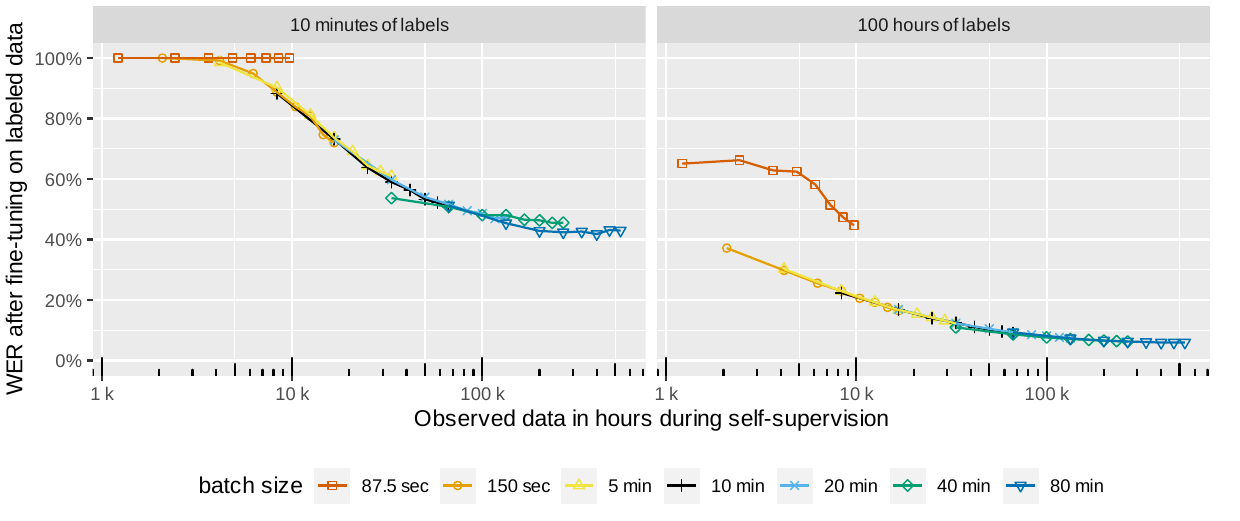}
    \caption{We plot the WER after fine-tuning against the hours of data processed during self-supervision (upper bound) for different batch sizes. The left column shows WER on LibriSpeech test-clean with 10 minutes of labeled data fine-tuning, the right column with 100 hours of labeled data fine-tuning.}
    \label{fig:ft-asr-time}
\end{figure*}

We will now focus on RQ~\ref{rq-equiv}. We compare the performance of batch sizes at different stages during pre-training. 
Because we use a cyclic learning rate, there are equivalences at each end of a cycle, at step $\{50\ \text{k},100\ \text{k}, \dots , 400\ \text{k}\}$. 
Different batches overlap on the amount data seen at particular checkpoints.
For example, 16.7\,k hours of data was observed with batch sizes of 150 sec, 5 min, 10 min, and 20 min respectively at 400\,k, 200\,k, 100\,k, and 50\,k steps. 
If less noisy gradient approximations are beneficial to learning, we expect a performance difference when we compare these checkpoints. 
However, if all that matters is observing more data, we expect no difference in performance between these checkpoints.

For each batch size, we fine-tune the checkpoints with an interval of 50\,k steps, resulting in 8 checkpoints per batch size.
We use the training methodology as described in \sec{asr-ft}.
For this experiment we focus on fine-tuning on 10 minutes and 100 hours of labeled data, with letter decoding, evaluating on the test-clean set. 
The results are shown in \fig{ft-asr-time}. 
We use the naive, upper bound of the amounts of data observed, as shown in \tab{epochs}.
We observe a direct relationship between the amount data observed during pre-traing and the WER after fine-tuning. 
There are only very minor differences between the WER of different checkpoints with the same amount of data. 
The curves for each batch size blend into each other, especially for the case of fine-tuning with 100 hours of data.
With 10 minutes of data we observe that a batch size of 40 minutes has slightly better performance at the start of training, and worse performance at the end of training, compared to batches of 20 minutes and 80 minutes. 
Also, we see that a batch size of 87.5 seconds is too small for good performance. 

\section{Discussion and conclusion} \label{sec:conclusion}

From the extensive search of batch sizes reported in \fig{ssl-metrics}, we see that in general larger batch sizes result in better downstream performance, if given the same amount of iteration during pre-training. 
This is consistent with the hypothesis of RQ \ref{rq-bs-pre} and \ref{rq-bs-down}.
We were surprised to observe convergence with a batch size of 150 sec, which appears to be the absolute minimum, as can be seen in the difference between the batch size of 87.5 and 150 sec in Figure \ref{fig:ssl-metrics}B, Figure \ref{fig:ft-asr-all} (10 min and 1 hour), and Figure \ref{fig:ft-asr-time}.  
A good indicator for well chosen hyperparameters is a continuous increase of the perplexity of the codebook logits (\fig{ssl-metrics}E-F).  With larger batch sizes, the similarity of codebook vectors decreases, which is an indication of the diversity of the learnt representations. 

In optimizing for LRs, we found the sub-linear heuristic perform well for all converging small batch sizes, and although we realize that an independent hyperparameter search for each batch size could improve performance a little, we are quite confident that this would not change our further conclusions.  

Regarding RC \ref{rq-bs-down} and \fig{ft-asr-all}, the observed performance follows expected behaviour, but we make this explicit for the first time, and find all data points, where relevant, in accordance with the original paper~\cite{wav2vec2_2020}, except where they decoded using a beam size of 500. Our cyclic LR schedule does not perform worse than \cite{wav2vec2_2020}, while allowing fine tuning experiments at regular intervals.  

It is remarkable, that even the diverging pre-training condition of 87.5\,sec (\fig{ssl-metrics}) show better fine-tuning performance than training from scratch in all test conditions (\fig{ft-asr-all}).  

The largest batch size we investigated showed a little worse performance, which we attribute to a need for stronger regularisation, e.g., dropout in more places or a higher value of~$\lambda_p$. 

In looking for an answer to why larger batch sizes are more effective, we saw in \fig{grad-var} that the standard deviation of the gradients reduces almost consistently with larger batch size, up to a critical value of 40\,min.  This is in fact consistent with the batch sizes reported in Hubert~\cite{hubert_2021} and Wavlm~\cite{wavlm_2022}.  


Regarding RC~\ref{rq-equiv}, we found that the most important factor for downstream task performance is the total amount of data seen during pre-training, i.e., the product of batch size and number of iterations, as shown convincingly in \fig{ft-asr-time}.  This means that it still is possible to carry out pretraining with limited amount of GPUs and/or memory, but one needs to be more patient or accept a penalty in performance, where \fig{ft-asr-all} can help in decision making.  

This brings us to the question if it benefits the community to benchmark SSL algorithms in speech by constraining the amount of data seen in training, e.g., to 100\,k hours.  In experiments with different algorithms, one might use 10\,k hours of seen data to reduce the computational burden, and verify conclusions at the 100\,k hours pre-training condition.  

Concluding, we observe that the batch size can be varied over a large range of values without a performance penalty, where the lower limit is set by convergence of the training (in our case 150 seconds), and the upper limit is one of diminishing returns~\cite{mccandlish2018empirical} (in our case 40 minutes).  These limits may be specific to the architecture (wav2vec 2.0 base, with approximately 95\,M parameters), but we believe that also larger models will show a dependence on the amount of data seen similar to \fig{ft-asr-time} in terms of the fine-tuning performance. 

\section*{Acknowledgments}
This work was carried out in part using the Snellius compute infrastructure under grant number 2022.022 from the Dutch Research Council NWO.

\section{References Section}
 
\bibliography{references}

\begin{thebibliography}{10}
\providecommand{\url}[1]{#1}
\csname url@samestyle\endcsname
\providecommand{\newblock}{\relax}
\providecommand{\bibinfo}[2]{#2}
\providecommand{\BIBentrySTDinterwordspacing}{\spaceskip=0pt\relax}
\providecommand{\BIBentryALTinterwordstretchfactor}{4}
\providecommand{\BIBentryALTinterwordspacing}{\spaceskip=\fontdimen2\font plus
\BIBentryALTinterwordstretchfactor\fontdimen3\font minus
  \fontdimen4\font\relax}
\providecommand{\BIBforeignlanguage}[2]{{%
\expandafter\ifx\csname l@#1\endcsname\relax
\typeout{** WARNING: IEEEtran.bst: No hyphenation pattern has been}%
\typeout{** loaded for the language `#1'. Using the pattern for}%
\typeout{** the default language instead.}%
\else
\language=\csname l@#1\endcsname
\fi
#2}}
\providecommand{\BIBdecl}{\relax}
\BIBdecl

\bibitem{wav2vec2_2020}
\BIBentryALTinterwordspacing
A.~Baevski, Y.~Zhou, A.~Mohamed, and M.~Auli, ``wav2vec 2.0: A framework for
  self-supervised learning of speech representations,'' in \emph{Advances in
  Neural Information Processing Systems}, vol.~33, 2020, pp. 12\,449--12\,460.
  [Online]. Available:
  \url{https://proceedings.neurips.cc/paper/2020/file/92d1e1eb1cd6f9fba3227870bb6d7f07-Paper.pdf}
\BIBentrySTDinterwordspacing

\bibitem{conneau_unsupervised_2021}
\BIBentryALTinterwordspacing
A.~Conneau, A.~Baevski, R.~Collobert, A.~Mohamed, and M.~Auli,
  ``\BIBforeignlanguage{en}{Unsupervised {Cross}-{Lingual} {Representation}
  {Learning} for {Speech} {Recognition}},'' in
  \emph{\BIBforeignlanguage{en}{Interspeech 2021}}.\hskip 1em plus 0.5em minus
  0.4em\relax ISCA, Aug. 2021, pp. 2426--2430. [Online]. Available:
  \url{https://www.isca-speech.org/archive/interspeech_2021/conneau21_interspeech.html}
\BIBentrySTDinterwordspacing

\bibitem{hubert_2021}
\BIBentryALTinterwordspacing
W.-N. Hsu, B.~Bolte, Y.-H.~H. Tsai, K.~Lakhotia, R.~Salakhutdinov, and
  A.~Mohamed, ``Hubert: Self-supervised speech representation learning by
  masked prediction of hidden units,'' \emph{IEEE/ACM Transactions on Audio,
  Speech, and Language Processing}, vol.~29, pp. 3451--3460, 2021. [Online].
  Available: \url{https://arxiv.org/abs/2106.07447}
\BIBentrySTDinterwordspacing

\bibitem{wavlm_2022}
\BIBentryALTinterwordspacing
S.~Chen, C.~Wang, Z.~Chen, Y.~Wu, S.~Liu, Z.~Chen, J.~Li, N.~Kanda,
  T.~Yoshioka, X.~Xiao, J.~Wu, L.~Zhou, S.~Ren, Y.~Qian, Y.~Qian, J.~Wu,
  M.~Zeng, X.~Yu, and F.~Wei, ``Wavlm: Large-scale self-supervised pre-training
  for full stack speech processing,'' \emph{IEEE Journal of Selected Topics in
  Signal Processing}, vol.~16, no.~6, pp. 1505--1518, 2022. [Online].
  Available: \url{https://arxiv.org/abs/2110.13900}
\BIBentrySTDinterwordspacing

\bibitem{yang21c_superb}
\BIBentryALTinterwordspacing
S.~wen Yang, P.-H. Chi, Y.-S. Chuang, C.-I.~J. Lai, K.~Lakhotia, Y.~Y. Lin,
  A.~T. Liu, J.~Shi, X.~Chang, G.-T. Lin, T.-H. Huang, W.-C. Tseng, K.~tik Lee,
  D.-R. Liu, Z.~Huang, S.~Dong, S.-W. Li, S.~Watanabe, A.~Mohamed, and
  H.~yi~Lee, ``{SUPERB: Speech Processing Universal PERformance Benchmark},''
  in \emph{Proc. Interspeech 2021}, 2021, pp. 1194--1198. [Online]. Available:
  \url{https://www.isca-speech.org/archive/interspeech_2021/yang21c_interspeech.html}
\BIBentrySTDinterwordspacing

\bibitem{whisper-radford23a}
\BIBentryALTinterwordspacing
A.~Radford, J.~W. Kim, T.~Xu, G.~Brockman, C.~Mcleavey, and I.~Sutskever,
  ``Robust speech recognition via large-scale weak supervision,'' in
  \emph{Proceedings of the 40th International Conference on Machine Learning},
  ser. Proceedings of Machine Learning Research, A.~Krause, E.~Brunskill,
  K.~Cho, B.~Engelhardt, S.~Sabato, and J.~Scarlett, Eds., vol. 202.\hskip 1em
  plus 0.5em minus 0.4em\relax PMLR, 23--29 Jul 2023, pp. 28\,492--28\,518.
  [Online]. Available: \url{https://proceedings.mlr.press/v202/radford23a.html}
\BIBentrySTDinterwordspacing

\bibitem{usm_zhang_google_2023}
\BIBentryALTinterwordspacing
Y.~Zhang, W.~Han, J.~Qin, Y.~Wang, A.~Bapna, Z.~Chen, N.~Chen, B.~Li,
  V.~Axelrod, G.~Wang, Z.~Meng, K.~Hu, A.~Rosenberg, R.~Prabhavalkar, D.~S.
  Park, P.~Haghani, J.~Riesa, G.~Perng, H.~Soltau, T.~Strohman, B.~Ramabhadran,
  T.~Sainath, P.~Moreno, C.-C. Chiu, J.~Schalkwyk, F.~Beaufays, and Y.~Wu,
  ``Google {USM}: {Scaling} {Automatic} {Speech} {Recognition} {Beyond} 100
  {Languages},'' Sep. 2023, arXiv:2303.01037 [cs, eess]. [Online]. Available:
  \url{http://arxiv.org/abs/2303.01037}
\BIBentrySTDinterwordspacing

\bibitem{panayotov_librispeech_2015}
\BIBentryALTinterwordspacing
V.~Panayotov, G.~Chen, D.~Povey, and S.~Khudanpur,
  ``\BIBforeignlanguage{en}{Librispeech: {An} {ASR} corpus based on public
  domain audio books},'' in \emph{\BIBforeignlanguage{en}{2015 {IEEE}
  {International} {Conference} on {Acoustics}, {Speech} and {Signal}
  {Processing} ({ICASSP})}}.\hskip 1em plus 0.5em minus 0.4em\relax South
  Brisbane, Queensland, Australia: IEEE, Apr. 2015, pp. 5206--5210. [Online].
  Available: \url{http://ieeexplore.ieee.org/document/7178964/}
\BIBentrySTDinterwordspacing

\bibitem{kahn2020libri}
\BIBentryALTinterwordspacing
J.~Kahn, M.~Rivi{\`e}re, W.~Zheng, E.~Kharitonov, Q.~Xu, P.-E. Mazar{\'e},
  J.~Karadayi, V.~Liptchinsky, R.~Collobert, C.~Fuegen \emph{et~al.},
  ``Libri-light: A benchmark for asr with limited or no supervision,'' in
  \emph{ICASSP 2020-2020 IEEE International Conference on Acoustics, Speech and
  Signal Processing (ICASSP)}.\hskip 1em plus 0.5em minus 0.4em\relax IEEE,
  2020, pp. 7669--7673. [Online]. Available:
  \url{https://arxiv.org/abs/1912.07875}
\BIBentrySTDinterwordspacing

\bibitem{chen21o_gigaspeech}
\BIBentryALTinterwordspacing
G.~Chen, S.~Chai, G.-B. Wang, J.~Du, W.-Q. Zhang, C.~Weng, D.~Su, D.~Povey,
  J.~Trmal, J.~Zhang, M.~Jin, S.~Khudanpur, S.~Watanabe, S.~Zhao, W.~Zou,
  X.~Li, X.~Yao, Y.~Wang, Z.~You, and Z.~Yan, ``{GigaSpeech: An Evolving,
  Multi-Domain ASR Corpus with 10,000 Hours of Transcribed Audio},'' in
  \emph{Proc. Interspeech 2021}, 2021, pp. 3670--3674. [Online]. Available:
  \url{https://www.isca-speech.org/archive/interspeech_2021/chen21o_interspeech.html}
\BIBentrySTDinterwordspacing

\bibitem{wang-etal-2021-voxpopuli}
\BIBentryALTinterwordspacing
C.~Wang, M.~Riviere, A.~Lee, A.~Wu, C.~Talnikar, D.~Haziza, M.~Williamson,
  J.~Pino, and E.~Dupoux, ``{V}ox{P}opuli: A large-scale multilingual speech
  corpus for representation learning, semi-supervised learning and
  interpretation,'' in \emph{Proceedings of the 59th Annual Meeting of the
  Association for Computational Linguistics and the 11th International Joint
  Conference on Natural Language Processing (Volume 1: Long Papers)}, C.~Zong,
  F.~Xia, W.~Li, and R.~Navigli, Eds.\hskip 1em plus 0.5em minus 0.4em\relax
  Online: Association for Computational Linguistics, Aug. 2021, pp. 993--1003.
  [Online]. Available: \url{https://aclanthology.org/2021.acl-long.80}
\BIBentrySTDinterwordspacing

\bibitem{mccandlish2018empirical}
\BIBentryALTinterwordspacing
S.~McCandlish, J.~Kaplan, D.~Amodei, and {OpenAI DotA team}, ``An empirical
  model of large-batch training,'' \emph{arXiv preprint arXiv:1812.06162},
  2018. [Online]. Available: \url{https://arxiv.org/abs/1812.06162}
\BIBentrySTDinterwordspacing

\bibitem{Shallue2018dataparallel}
\BIBentryALTinterwordspacing
C.~J. Shallue, J.~Lee, J.~Antognini, J.~Sohl-Dickstein, R.~Frostig, and G.~E.
  Dahl, ``Measuring the effects of data parallelism on neural network
  training,'' \emph{Journal of Machine Learning Research}, vol.~20, no. 112,
  pp. 1--49, 2019. [Online]. Available:
  \url{http://jmlr.org/papers/v20/18-789.html}
\BIBentrySTDinterwordspacing

\bibitem{goodfellow2016deep}
\BIBentryALTinterwordspacing
I.~Goodfellow, Y.~Bengio, and A.~Courville, \emph{Deep learning {(chapter
  15)}}.\hskip 1em plus 0.5em minus 0.4em\relax MIT press, 2016. [Online].
  Available:
  \url{https://www.deeplearningbook.org/contents/representation.html}
\BIBentrySTDinterwordspacing

\bibitem{mohamed2022self}
\BIBentryALTinterwordspacing
A.~Mohamed, H.-y. Lee, L.~Borgholt, J.~D. Havtorn, J.~Edin, C.~Igel,
  K.~Kirchhoff, S.-W. Li, K.~Livescu, L.~Maal{\o}e \emph{et~al.},
  ``Self-supervised speech representation learning: A review,'' \emph{IEEE
  Journal of Selected Topics in Signal Processing}, 2022. [Online]. Available:
  \url{https://arxiv.org/abs/2205.10643}
\BIBentrySTDinterwordspacing

\bibitem{van2017neural}
\BIBentryALTinterwordspacing
A.~Van Den~Oord, O.~Vinyals \emph{et~al.}, ``Neural discrete representation
  learning,'' \emph{Advances in neural information processing systems},
  vol.~30, 2017. [Online]. Available:
  \url{https://proceedings.neurips.cc/paper/2017/hash/7a98af17e63a0ac09ce2e96d03992fbc-Abstract.html}
\BIBentrySTDinterwordspacing

\bibitem{liu2020mockingjay}
\BIBentryALTinterwordspacing
A.~T. Liu, S.-w. Yang, P.-H. Chi, P.-c. Hsu, and H.-y. Lee, ``Mockingjay:
  Unsupervised speech representation learning with deep bidirectional
  transformer encoders,'' in \emph{ICASSP 2020-2020 IEEE International
  Conference on Acoustics, Speech and Signal Processing (ICASSP)}.\hskip 1em
  plus 0.5em minus 0.4em\relax IEEE, 2020, pp. 6419--6423. [Online]. Available:
  \url{https://arxiv.org/abs/1910.12638}
\BIBentrySTDinterwordspacing

\bibitem{ling2020deep}
\BIBentryALTinterwordspacing
S.~Ling, Y.~Liu, J.~Salazar, and K.~Kirchhoff, ``Deep contextualized acoustic
  representations for semi-supervised speech recognition,'' in \emph{ICASSP
  2020-2020 IEEE International Conference on Acoustics, Speech and Signal
  Processing (ICASSP)}.\hskip 1em plus 0.5em minus 0.4em\relax IEEE, 2020, pp.
  6429--6433. [Online]. Available: \url{https://arxiv.org/abs/1912.01679}
\BIBentrySTDinterwordspacing

\bibitem{ling2020decoar2}
\BIBentryALTinterwordspacing
S.~Ling and Y.~Liu, ``Decoar 2.0: Deep contextualized acoustic representations
  with vector quantization,'' \emph{arXiv preprint arXiv:2012.06659}, 2020.
  [Online]. Available: \url{https://arxiv.org/abs/2012.06659}
\BIBentrySTDinterwordspacing

\bibitem{liu2021tera}
\BIBentryALTinterwordspacing
A.~T. Liu, S.-W. Li, and H.-y. Lee, ``Tera: Self-supervised learning of
  transformer encoder representation for speech,'' \emph{IEEE/ACM Transactions
  on Audio, Speech, and Language Processing}, vol.~29, pp. 2351--2366, 2021.
  [Online]. Available: \url{https://arxiv.org/abs/2007.06028}
\BIBentrySTDinterwordspacing

\bibitem{milde18_interspeech}
\BIBentryALTinterwordspacing
B.~Milde and C.~Biemann, ``{Unspeech: Unsupervised Speech Context
  Embeddings},'' in \emph{Proc. Interspeech 2018}, 2018, pp. 2693--2697.
  [Online]. Available:
  \url{https://www.isca-speech.org/archive/interspeech_2018/milde18_interspeech.html}
\BIBentrySTDinterwordspacing

\bibitem{oord2018representation}
\BIBentryALTinterwordspacing
A.~v.~d. Oord, Y.~Li, and O.~Vinyals, ``Representation learning with
  contrastive predictive coding,'' \emph{arXiv preprint arXiv:1807.03748},
  2018. [Online]. Available: \url{https://arxiv.org/abs/1807.03748}
\BIBentrySTDinterwordspacing

\bibitem{schneider19_interspeech}
\BIBentryALTinterwordspacing
S.~Schneider, A.~Baevski, R.~Collobert, and M.~Auli, ``{wav2vec: Unsupervised
  Pre-Training for Speech Recognition},'' in \emph{Proc. Interspeech 2019},
  2019, pp. 3465--3469. [Online]. Available:
  \url{https://www.isca-speech.org/archive/interspeech_2019/schneider19_interspeech.html}
\BIBentrySTDinterwordspacing

\bibitem{baevskivqwav2vec}
\BIBentryALTinterwordspacing
A.~Baevski, S.~Schneider, and M.~Auli, ``vq-wav2vec: Self-supervised learning
  of discrete speech representations,'' in \emph{8th International Conference
  on Learning Representations, {ICLR} 2020, Addis Ababa, Ethiopia, April 26-30,
  2020}.\hskip 1em plus 0.5em minus 0.4em\relax OpenReview.net, 2020. [Online].
  Available: \url{https://openreview.net/forum?id=rylwJxrYDS}
\BIBentrySTDinterwordspacing

\bibitem{sadhu21_interspeech}
\BIBentryALTinterwordspacing
S.~Sadhu, D.~He, C.-W. Huang, S.~H. Mallidi, M.~Wu, A.~Rastrow, A.~Stolcke,
  J.~Droppo, and R.~Maas, ``{wav2vec-C: A Self-Supervised Model for Speech
  Representation Learning},'' in \emph{Proc. Interspeech 2021}, 2021, pp.
  711--715. [Online]. Available:
  \url{https://www.isca-speech.org/archive/interspeech_2021/sadhu21_interspeech.html}
\BIBentrySTDinterwordspacing

\bibitem{baevski2022data2vec}
\BIBentryALTinterwordspacing
A.~Baevski, W.-N. Hsu, Q.~Xu, A.~Babu, J.~Gu, and M.~Auli, ``Data2vec: A
  general framework for self-supervised learning in speech, vision and
  language,'' in \emph{International Conference on Machine Learning}.\hskip 1em
  plus 0.5em minus 0.4em\relax PMLR, 2022, pp. 1298--1312. [Online]. Available:
  \url{https://proceedings.mlr.press/v162/baevski22a.html}
\BIBentrySTDinterwordspacing

\bibitem{liu2023dinosr}
A.~H. Liu, H.-J. Chang, M.~Auli, W.-N. Hsu, and J.~R. Glass, ``Dinosr:
  Self-distillation and online clustering for self-supervised speech
  representation learning,'' \emph{arXiv preprint arXiv:2305.10005}, 2023.

\bibitem{kaplan2020scaling}
\BIBentryALTinterwordspacing
J.~Kaplan, S.~McCandlish, T.~Henighan, T.~B. Brown, B.~Chess, R.~Child,
  S.~Gray, A.~Radford, J.~Wu, and D.~Amodei, ``Scaling laws for neural language
  models,'' \emph{arXiv preprint arXiv:2001.08361}, 2020. [Online]. Available:
  \url{https://arxiv.org/abs/2001.08361}
\BIBentrySTDinterwordspacing

\bibitem{goyal2019scaling}
\BIBentryALTinterwordspacing
P.~Goyal, D.~Mahajan, A.~Gupta, and I.~Misra, ``Scaling and benchmarking
  self-supervised visual representation learning,'' in \emph{Proceedings of the
  ieee/cvf International Conference on computer vision}, 2019, pp. 6391--6400.
  [Online]. Available:
  \url{https://openaccess.thecvf.com/content_ICCV_2019/papers/Goyal_Scaling_and_Benchmarking_Self-Supervised_Visual_Representation_Learning_ICCV_2019_paper.pdf}
\BIBentrySTDinterwordspacing

\bibitem{pu21_interspeech}
\BIBentryALTinterwordspacing
J.~Pu, Y.~Yang, R.~Li, O.~Elibol, and J.~Droppo, ``{Scaling Effect of
  Self-Supervised Speech Models},'' in \emph{Proc. Interspeech 2021}, 2021, pp.
  1084--1088. [Online]. Available:
  \url{https://www.isca-speech.org/archive/interspeech_2021/pu21_interspeech.html}
\BIBentrySTDinterwordspacing

\bibitem{pmlr-v119-chen20j}
\BIBentryALTinterwordspacing
T.~Chen, S.~Kornblith, M.~Norouzi, and G.~Hinton, ``A simple framework for
  contrastive learning of visual representations,'' in \emph{Proceedings of the
  37th International Conference on Machine Learning}, ser. Proceedings of
  Machine Learning Research, H.~D. III and A.~Singh, Eds., vol. 119.\hskip 1em
  plus 0.5em minus 0.4em\relax PMLR, 13--18 Jul 2020, pp. 1597--1607. [Online].
  Available: \url{https://proceedings.mlr.press/v119/chen20j.html}
\BIBentrySTDinterwordspacing

\bibitem{pmlr-v139-radford21a}
\BIBentryALTinterwordspacing
A.~Radford, J.~W. Kim, C.~Hallacy, A.~Ramesh, G.~Goh, S.~Agarwal, G.~Sastry,
  A.~Askell, P.~Mishkin, J.~Clark, G.~Krueger, and I.~Sutskever, ``Learning
  transferable visual models from natural language supervision,'' in
  \emph{Proceedings of the 38th International Conference on Machine Learning},
  ser. Proceedings of Machine Learning Research, M.~Meila and T.~Zhang, Eds.,
  vol. 139.\hskip 1em plus 0.5em minus 0.4em\relax PMLR, 18--24 Jul 2021, pp.
  8748--8763. [Online]. Available:
  \url{https://proceedings.mlr.press/v139/radford21a.html}
\BIBentrySTDinterwordspacing

\bibitem{yuan2021florence}
\BIBentryALTinterwordspacing
L.~Yuan, D.~Chen, Y.-L. Chen, N.~Codella, X.~Dai, J.~Gao, H.~Hu, X.~Huang,
  B.~Li, C.~Li \emph{et~al.}, ``Florence: A new foundation model for computer
  vision,'' \emph{arXiv preprint arXiv:2111.11432}, 2021. [Online]. Available:
  \url{https://arxiv.org/abs/2111.11432}
\BIBentrySTDinterwordspacing

\bibitem{pmlr-v137-mitrovic20a}
\BIBentryALTinterwordspacing
J.~Mitrovic, B.~McWilliams, and M.~Rey, ``Less can be more in contrastive
  learning,'' in \emph{Proceedings on "I Can't Believe It's Not Better!" at
  NeurIPS Workshops}, ser. Proceedings of Machine Learning Research,
  J.~Zosa~Forde, F.~Ruiz, M.~F. Pradier, and A.~Schein, Eds., vol. 137.\hskip
  1em plus 0.5em minus 0.4em\relax PMLR, 12 Dec 2020, pp. 70--75. [Online].
  Available: \url{https://proceedings.mlr.press/v137/mitrovic20a.html}
\BIBentrySTDinterwordspacing

\bibitem{Chen2022ClGradBias}
\BIBentryALTinterwordspacing
C.~Chen, J.~Zhang, Y.~Xu, L.~Chen, J.~Duan, Y.~Chen, S.~Tran, B.~Zeng, and
  T.~Chilimbi, ``Why do we need large batchsizes in contrastive learning? a
  gradient-bias perspective,'' in \emph{Advances in Neural Information
  Processing Systems}, vol.~35.\hskip 1em plus 0.5em minus 0.4em\relax Curran
  Associates, Inc., 2022, pp. 33\,860--33\,875. [Online]. Available:
  \url{https://proceedings.neurips.cc/paper_files/paper/2022/file/db174d373133dcc6bf83bc98e4b681f8-Paper-Conference.pdf}
\BIBentrySTDinterwordspacing

\bibitem{izsak-etal-2021-train}
\BIBentryALTinterwordspacing
P.~Izsak, M.~Berchansky, and O.~Levy, ``How to train {BERT} with an academic
  budget,'' in \emph{Proceedings of the 2021 Conference on Empirical Methods in
  Natural Language Processing}, Nov. 2021, pp. 10\,644--10\,652. [Online].
  Available: \url{https://aclanthology.org/2021.emnlp-main.831}
\BIBentrySTDinterwordspacing

\bibitem{devlin-etal-2019-bert}
\BIBentryALTinterwordspacing
J.~Devlin, M.-W. Chang, K.~Lee, and K.~Toutanova, ``{BERT}: Pre-training of
  deep bidirectional transformers for language understanding,'' in
  \emph{Proceedings of the 2019 Conference of the North {A}merican Chapter of
  the Association for Computational Linguistics: Human Language Technologies,
  Volume 1 (Long and Short Papers)}, Jun. 2019, pp. 4171--4186. [Online].
  Available: \url{https://aclanthology.org/N19-1423}
\BIBentrySTDinterwordspacing

\bibitem{rajbhandari2020zero}
\BIBentryALTinterwordspacing
S.~Rajbhandari, J.~Rasley, O.~Ruwase, and Y.~He, ``Zero: Memory optimizations
  toward training trillion parameter models,'' in \emph{SC20: International
  Conference for High Performance Computing, Networking, Storage and
  Analysis}.\hskip 1em plus 0.5em minus 0.4em\relax IEEE, 2020, pp. 1--16.
  [Online]. Available: \url{https://arxiv.org/abs/1910.02054}
\BIBentrySTDinterwordspacing

\bibitem{MicikeviciusNAD18}
\BIBentryALTinterwordspacing
P.~Micikevicius, S.~Narang, J.~Alben, G.~F. Diamos, E.~Elsen, D.~Garc{\'{\i}}a,
  B.~Ginsburg, M.~Houston, O.~Kuchaiev, G.~Venkatesh, and H.~Wu, ``Mixed
  precision training,'' in \emph{6th International Conference on Learning
  Representations, {ICLR} 2018}, 2018. [Online]. Available:
  \url{https://openreview.net/forum?id=r1gs9JgRZ}
\BIBentrySTDinterwordspacing

\bibitem{chen23l_interspeech}
\BIBentryALTinterwordspacing
W.~Chen, X.~Chang, Y.~Peng, Z.~Ni, S.~Maiti, and S.~Watanabe, ``{Reducing
  Barriers to Self-Supervised Learning: HuBERT Pre-training with Academic
  Compute},'' in \emph{Proc. INTERSPEECH 2023}, 2023, pp. 4404--4408. [Online].
  Available:
  \url{https://www.isca-speech.org/archive/pdfs/interspeech_2023/chen23l_interspeech.pdf}
\BIBentrySTDinterwordspacing

\bibitem{S3L}
\BIBentryALTinterwordspacing
Y.-H. Cao and J.~Wu, ``Rethinking self-supervised learning: Small is
  beautiful,'' \emph{arXiv preprint arXiv:2103.13559}, 2021. [Online].
  Available: \url{https://arxiv.org/abs/2103.13559}
\BIBentrySTDinterwordspacing

\bibitem{wu2018groupnorm}
\BIBentryALTinterwordspacing
Y.~Wu and K.~He, ``Group normalization,'' in \emph{Proceedings of the European
  Conference on Computer Vision (ECCV)}, September 2018. [Online]. Available:
  \url{https://arxiv.org/abs/1803.08494}
\BIBentrySTDinterwordspacing

\bibitem{ba2016layernorm}
J.~L. Ba, J.~R. Kiros, and G.~E. Hinton, ``Layer normalization,'' \emph{arXiv
  preprint arXiv:1607.06450}, 2016.

\bibitem{salimans2016weight}
\BIBentryALTinterwordspacing
T.~Salimans and D.~P. Kingma, ``Weight normalization: A simple
  reparameterization to accelerate training of deep neural networks,'' in
  \emph{Advances in Neural Information Processing Systems}, vol.~29, 2016.
  [Online]. Available:
  \url{https://proceedings.neurips.cc/paper_files/paper/2016/file/ed265bc903a5a097f61d3ec064d96d2e-Paper.pdf}
\BIBentrySTDinterwordspacing

\bibitem{jang2017categorical}
\BIBentryALTinterwordspacing
E.~Jang, S.~Gu, and B.~Poole, ``Categorical reparameterization with
  gumbel-softmax,'' in \emph{5th International Conference on Learning
  Representations, {ICLR} 2017, Toulon, France, April 24-26, 2017, Conference
  Track Proceedings}, 2017. [Online]. Available:
  \url{https://openreview.net/forum?id=rkE3y85ee}
\BIBentrySTDinterwordspacing

\bibitem{park19e_specaugment}
D.~S. Park, W.~Chan, Y.~Zhang, C.-C. Chiu, B.~Zoph, E.~D. Cubuk, and Q.~V. Le,
  ``{SpecAugment: A Simple Data Augmentation Method for Automatic Speech
  Recognition},'' in \emph{Proc. Interspeech 2019}, 2019, pp. 2613--2617.

\bibitem{Graves:2006}
\BIBentryALTinterwordspacing
A.~Graves, S.~Fern{\'a}ndez, F.~Gomez, and J.~Schmidhuber, ``Connectionist
  temporal classification: labelling unsegmented sequence data with recurrent
  neural networks,'' in \emph{Proceedings of the 23rd international conference
  on Machine learning}, 2006, pp. 369--376. [Online]. Available:
  \url{https://archive.air.in.tum.de/Main/Publications/Graves2006a.pdf}
\BIBentrySTDinterwordspacing

\end{thebibliography}
\bibliographystyle{IEEEtran}

{\appendices
\section*{Fine-tuning best checkpoint for each batch size in tabular format}

In Table \ref{tab:ft-asr-all} we show the data in \fig{ft-asr-all} in tabular format. 

\begin{table*}[b]
\centering
\caption{The WER after fine-tuning the best SSL checkpoint for each batch size. Fine-tuning is done on 5 different amounts of label conditions. Decoding is done with letter decoding as well as 4-gram word decoding. }
\label{tab:ft-asr-all}
\begin{tabular}{@{}lclllllll@{}}
\toprule
             &                &  & \multicolumn{2}{c}{letter decoding} &  & \multicolumn{2}{c}{4-gram word decoding} &  \\ \midrule
labeled data & SSL batch size &  & test-clean       & test-other       &  & test-clean          & test-other         &  \\ \midrule
10 min       & scratch        &  & 129.45           & 129.47           &  & 111.17              & 111.95             &  \\
             & 87.5 sec       &  & 101.77           & 102.07           &  & 96.8                & 97.34              &  \\
             & 150 sec        &  & 72.07            & 82.02            &  & 54.44               & 68.26              &  \\
             & 5 min          &  & 61.26            & 71.06            &  & 40.96               & 53.53              &  \\
             & 10 min         &  & 50.39            & 59.45            &  & 32.06               & 42.32              &  \\
             & 20 min         &  & 47.04            & 54.67            &  & 27.3                & 35.96              &  \\
             & 40 min         &  & 45.44            & 52.05            &  & 26.89               & 34.35              &  \\
             & 80 min         &  & 41.64            & 50.06            &  & 24.38               & 34.47              &  \\ \midrule
1 hour       & scratch        &  & 106.68           & 104.76           &  & 101.23              & 100.09             &  \\
             & 87.5 sec       &  & 102.21           & 103.06           &  & 93.72               & 95.61              &  \\
             & 150 sec        &  & 53.28            & 68.37            &  & 34.00               & 52.00              &  \\
             & 5 min          &  & 39.04            & 53.57            &  & 22.31               & 37.11              &  \\
             & 10 min         &  & 28.83            & 41.47            &  & 15.69               & 27.03              &  \\
             & 20 min         &  & 25.12            & 35.23            &  & 13.14               & 22.25              &  \\
             & 40 min         &  & 23.26            & 31.75            &  & 11.90               & 19.52              &  \\
             & 80 min         &  & 21.94            & 31.30            &  & 11.80               & 20.72              &  \\ \midrule
10 hours     & scratch        &  & 105.59           & 104.03           &  & 100.56              & 99.25              &  \\
             & 87.5 sec       &  & 92.21            & 98.09            &  & 78.50               & 89.18              &  \\
             & 150 sec        &  & 34.53            & 54.43            &  & 20.20               & 39.96              &  \\
             & 5 min          &  & 23.24            & 40.24            &  & 13.42               & 28.07              &  \\
             & 10 min         &  & 16.27            & 29.51            &  & 9.46                & 20.27              &  \\
             & 20 min         &  & 13.26            & 23.83            &  & 7.66                & 16.31              &  \\
             & 40 min         &  & 11.11            & 19.78            &  & 6.59                & 13.46              &  \\
             & 80 min         &  & 10.38            & 19.56            &  & 6.28                & 13.67              &  \\ \midrule
100 hours    & scratch        &  & 69.02            & 84.70            &  & 50.36               & 73.44              &  \\
             & 87.5 sec       &  & 44.74            & 70.44            &  & 26.41               & 55.58              &  \\
             & 150 sec        &  & 16.68            & 40.41            &  & 9.93                & 29.63              &  \\
             & 5 min          &  & 12.38            & 31.43            &  & 7.72                & 22.66              &  \\
             & 10 min         &  & 9.10             & 23.43            &  & 5.93                & 16.87              &  \\
             & 20 min         &  & 7.41             & 18.93            &  & 5.12                & 13.72              &  \\
             & 40 min         &  & 6.24             & 14.87            &  & 4.38                & 10.88              &  \\
             & 80 min         &  & 5.86             & 15.97            &  & 4.45                & 12.05              &  \\ \midrule
960 hours    & scratch        &  & 25.45            & 48.65            &  & 12.62               & 32.20              &  \\
             & 87.5 sec       &  & 12.74            & 31.39            &  & 6.83                & 20.84              &  \\
             & 150 sec        &  & 7.05             & 19.60            &  & 4.35                & 12.77              &  \\
             & 5 min          &  & 6.01             & 16.84            &  & 4.12                & 11.45              &  \\
             & 10 min         &  & 5.27             & 13.91            &  & 3.74                & 9.66               &  \\
             & 20 min         &  & 4.44             & 12.53            &  & 3.33                & 8.90               &  \\
             & 40 min         &  & 4.12             & 10.62            &  & 3.20                & 7.87               &  \\
             & 80 min         &  & 4.22             & 11.58            &  & 3.34                & 8.68               &  \\ \bottomrule
\end{tabular}
\end{table*}

\section*{Performance with overlapping amounts of observed data in tabular format}

In Table \ref{tab:overlap-hours-seen} we show the data in \fig{ft-asr-time} in tabular format, where at least 2 batch sizes match exactly in the amount of data seen during self-supervision. 

\begin{table*}[t]
\centering
\caption{The WER after fine-tuning SSL checkpoints with overlapping amount of hour seen during self-supervision, but using a different batch size and number of iterations.}
\label{tab:overlap-hours-seen}
\begin{tabular}{@{}rrrrrrrrrl@{}}
\toprule
\multicolumn{3}{c}{during self-supervised learning} & \multicolumn{1}{c}{} & \multicolumn{2}{c}{\begin{tabular}[c]{@{}c@{}}fine-tuning with \\ 10 minutes of labels \\ (WER in \%)\end{tabular}} & \multicolumn{1}{c}{} & \multicolumn{2}{c}{\begin{tabular}[c]{@{}c@{}}fine-tuning with\\ 100 hours of labels \\ (WER in \%)\end{tabular}} &  \\ \midrule
hours seen       & batch size      & iteration      &                      & test-clean                                               & test-other                                               &                      & test-clean                                              & test-other                                              &  \\ \midrule
4.17 k           & 150 sec         & 100 k          &                      & 99.16                                                    & 101.33                                                   &                      & 29.79                                                   & 57.52                                                   &  \\
4.17 k           & 5 min           & 50 k           &                      & 98.55                                                    & 100.04                                                   &                      & 30.25                                                   & 58.14                                                   &  \\ \midrule
8.33 k           & 150 sec         & 200 k          &                      & 88.71                                                    & 94.32                                                    &                      & 22.98                                                   & 50.31                                                   &  \\
8.33 k           & 5 min           & 100 k          &                      & 90.23                                                    & 95.29                                                    &                      & 22.98                                                   & 50.57                                                   &  \\
8.33 k           & 10 min          & 50 k           &                      & 88.35                                                    & 93.96                                                    &                      & 22.35                                                   & 49.56                                                   &  \\ \midrule
12.5 k           & 150 sec         & 300 k          &                      & 80.83                                                    & 88.28                                                    &                      & 19.23                                                   & 44.77                                                   &  \\
12.5 k           & 5 min           & 150 k          &                      & 81.17                                                    & 89.71                                                    &                      & 19.27                                                   & 45.25                                                   &  \\ \midrule
16.7 k           & 150 sec         & 400 k          &                      & 72.00                                                    & 82.15                                                    &                      & 16.49                                                   & 40.18                                                   &  \\
16.7 k           & 5 min           & 200 k          &                      & 73.63                                                    & 83.54                                                    &                      & 16.71                                                   & 41.17                                                   &  \\
16.7 k           & 10 min          & 100 k          &                      & 73.17                                                    & 84.01                                                    &                      & 16.83                                                   & 40.63                                                   &  \\
16.7 k           & 20 min          & 50 k           &                      & 72.95                                                    & 83.12                                                    &                      & 16.89                                                   & 41.09                                                   &  \\ \midrule
25 k             & 5 min           & 300 k          &                      & 64.40                                                    & 75.36                                                    &                      & 14.21                                                   & 35.51                                                   &  \\
25 k             & 10 min          & 150 k          &                      & 63.73                                                    & 74.55                                                    &                      & 13.97                                                   & 35.50                                                   &  \\ \midrule
33.3 k           & 5 min           & 400 k          &                      & 60.96                                                    & 70.61                                                    &                      & 12.24                                                   & 31.49                                                   &  \\
33.3 k           & 10 min          & 200 k          &                      & 59.00                                                    & 69.44                                                    &                      & 12.46                                                   & 31.39                                                   &  \\
33.3 k           & 20 min          & 100 k          &                      & 59.76                                                    & 70.14                                                    &                      & 12.30                                                   & 31.22                                                   &  \\
33.3 k           & 40 min          & 50 k           &                      & 53.70                                                    & 63.52                                                    &                      & 10.94                                                   & 27.53                                                   &  \\ \midrule
50 k             & 10 min          & 300 k          &                      & 53.32                                                    & 63.30                                                    &                      & 10.21                                                   & 25.99                                                   &  \\
50 k             & 20 min          & 150 k          &                      & 54.04                                                    & 63.91                                                    &                      & 10.47                                                   & 26.55                                                   &  \\ \midrule
66.7 k           & 10 min          & 400 k          &                      & 50.71                                                    & 60.05                                                    &                      & 9.07                                                    & 23.06                                                   &  \\
66.7 k           & 20 min          & 200 k          &                      & 51.64                                                    & 60.64                                                    &                      & 9.28                                                    & 23.97                                                   &  \\
66.7 k           & 40 min          & 100 k          &                      & 50.80                                                    & 58.74                                                    &                      & 8.63                                                    & 21.26                                                   &  \\
66.7 k           & 80 min          & 50 k           &                      & 51.26                                                    & 60.42                                                    &                      & 9.29                                                    & 23.20                                                   &  \\ \midrule
100 k            & 20 min          & 300 k          &                      & 48.61                                                    & 56.51                                                    &                      & 8.09                                                    & 20.75                                                   &  \\
100 k            & 40 min          & 150 k          &                      & 48.04                                                    & 55.41                                                    &                      & 7.63                                                    & 18.67                                                   &  \\ \midrule
133 k            & 20 min          & 400 k          &                      & 47.43                                                    & 54.71                                                    &                      & 7.40                                                    & 19.04                                                   &  \\
133 k            & 40 min          & 200 k          &                      & 48.06                                                    & 55.66                                                    &                      & 7.11                                                    & 17.25                                                   &  \\
133 k            & 80 min          & 100 k          &                      & 45.40                                                    & 53.46                                                    &                      & 7.31                                                    & 18.32                                                   &  \\ \midrule
200 k            & 40 min          & 300 k          &                      & 46.35                                                    & 53.42                                                    &                      & 6.58                                                    & 15.74                                                   &  \\
200 k            & 80 min          & 150 k          &                      & 42.84                                                    & 51.20                                                    &                      & 6.58                                                    & 16.70                                                   &  \\ \midrule
267 k            & 40 min          & 400 k          &                      & 45.57                                                    & 52.50                                                    &                      & 6.26                                                    & 15.10                                                   &  \\
267 k            & 80 min          & 200 k          &                      & 42.40                                                    & 50.34                                                    &                      & 6.27                                                    & 15.85                                                   &  \\ \bottomrule
\end{tabular}
\end{table*}

}

\end{document}